\documentclass[12pt]{amsart}

\usepackage{amssymb}
\usepackage[centertags]{amsmath}

\newenvironment{prf}{\noindent{\it Proof\/}:}{$\;\square$\par\medskip}
\newenvironment{example}
{\noindent{\bf Example\/}:}{\par}
\newtheorem%
{thm}{Theorem}[section]
\newtheorem%
{proposition}[thm]{Proposition}
\newtheorem%
{lemma}[thm]{Lemma}
\newtheorem%
{definition}[thm]{Definition}
\newtheorem%
{corollary}[thm]{Corollary}
\newtheorem%
{conjecture}[thm]{Conjecture} \theoremstyle{definition}

\newtheorem{defn}{Definition}[section]
\theoremstyle{remark}
\newtheorem{rem}{Remark}[section]

\newcommand{\dontprint}[1]{\relax}

\def\gh{\mathrm{gh}}
\title
[]
{Quantizing non-Lagrangian gauge theories: \\
an augmentation method}
\author{S.L. Lyakhovich and  A.A. Sharapov}
\address{Department of Quantum Field Theory, Tomsk State University, Tomsk 634050, Russia}
\email{sll@phys.tsu.ru, sharapov@phys.tsu.ru}

\thanks{We are thankful to G.~Barnich and
M.~Henneaux for useful discussions of this work, and we appreciate
warm hospitality of International Solvay Institutes and Department
of Mathematical and Theoretical Physics of University Libr\'e de
Bruxelles where this work has been finalized. The work was
partially supported by the RFBR grant 06-02-17352 and the grant
for Support of Russian Scientific Schools 5103.2006.2 and also
benefited from the ``Innovative Education Programme'' of TSU}

\begin{document}

\begin{abstract}
We discuss a recently proposed method of quantizing general
non-Lagrangian gauge theories. The method can be implemented in
many different ways, in particular, it can employ a conversion
procedure that turns an original non-Lagrangian field theory in
$d$ dimensions into an equivalent Lagrangian, topological field
theory in $d+1$ dimensions. The method involves, besides the
classical equations of motion, one more geometric ingredient
called the Lagrange anchor. Different Lagrange anchors result in
different quantizations of one and the same classical theory.
Given the classical equations of motion and Lagrange anchor as
input data, a new procedure, called the augmentation, is proposed
to quantize non-Lagrangian dynamics. Within the augmentation
procedure, the originally non-Lagrangian theory is absorbed by a
wider Lagrangian theory on the same space-time manifold. The
augmented theory is not generally equivalent to the original one
as it has more physical degrees of freedom than the original
theory. However, the extra degrees of freedom are factorized out
in a certain regular way both at classical and quantum levels. The
general techniques are exemplified by quantizing two
non-Lagrangian models of physical interest.
\end{abstract}

\maketitle

\newpage

\tableofcontents

\newpage

\section{Introduction}
Classical dynamics can be consistently formulated in terms of
equations of motion alone. The variational principle, being a
useful tool for studying various aspects  of classical dynamics,
is not needed to have the classical theory defined as such.
However, to promote the classical dynamics to quantum level, it is
insufficient to know only the equations of motion, one or another
extra structure is needed. If the quantum theory is supposed to be
formulated  in the language  of  Feynman's path integrals, it is
the action functional that can serve as the additional ingredient
needed for quantization. On the other hand, any Lagrangian
equations of motion can always be brought to a (constrained)
Hamiltonian form that makes possible applying canonical
quantization. Furthermore, the method of deformation quantization
applies to the Hamiltonian systems even though the underlying
Poisson bracket is degenerate \cite{Ko} (in which case the
Hamilton equations can have no variational formulation). As it has
been recently found \cite{LS1}, \cite{CaFe1}, the deformation
quantization can also be implemented under far less restrictive
conditions on the equations of motion than the requirement to be
Hamiltonian. Roughly speaking, the phase-space evolution flow is
not required to be Hamiltonian: It is sufficient if the evolution
preserves the Poisson bracket modulo constraints and gauge
transformations. The bracket, in its turn,  is also required to
satisfy the Jacobi identity in a weak sense, i.e., modulo
constraints and gauge transformations. For accurate definitions,
see \cite{LS1}, \cite{CaFe1}.

So, the deformation quantization has progressed in recent years
reaching far beyond the range of theories admitting variational
principle for equations of motion. At the same time, the methods
of constructing the partition functions\footnote{In the following
we will also use the term \textit{probability amplitude} as a
synonym for the \textit{partition function}.} dating back to
Feynman, Schwinger and Dyson, and being now developed in full
generality for arbitrary Lagrangian gauge theories \cite{DW},
\cite{BV}, \cite{HT}, have not made much progress in the class of
theories having no action functional. Until recently, no general
method has been known to path-integral quantize a non-Lagrangian
theory as it was not clear what might be a generalization of the
familiar Schwinger-Dyson's equation in the situation where no
Lagrangian formulation is possible for the classical equations of
motion.

In our recent papers \cite{KLS}, \cite{LS2}, we have identified a
general structure, called the Lagrange anchor, which is
determinative for the quantization in terms of partition functions
in the same sense as the Poisson bracket defines deformation
quantization in terms of a star product. The Lagrange anchor is a
geometric object that can be interpreted in many different ways.
In particular, one could say that the Lagrange anchor is related
to the canonical anti-bracket of the Batalin-Vilkovisky formalism
\cite{BV} much like a generic (i.e., possibly degenerate and
non-constant rank) Poisson bracket is related to the canonical
Poisson bracket. The anchor is also required to satisfy certain
compatibility conditions involving equations of motion. In
Lagrangian theory, these conditions are automatically satisfied
for the canonical anti-bracket in consequence of the fact that the
equations of motion are variations of the action functional. If
the anchor is invertible, these compatibility conditions ensure
existence of the equivalent Lagrangian formulation. It turns out
that the partition function can still be constructed by making use
only of the equations of motion and the Lagrange anchor, even
though the latter is degenerate, defining no action functional.
The next section contains an accurate definition of the Lagrange
anchor and discussion of its properties.

By now two procedures have been worked out to construct partition
functions for general non-Lagrangian theories. The first one
\cite{KLS} suggests a conversion of the original non-Lagrangian
field theory in $d$ dimensions into an equivalent
$(d+1)$-dimensional Lagrangian topological theory which can then
be quantized by the standard BV method. The conversion procedure
is quite ambiguous and essentially depends on the choice of the
Lagrange anchor.  If the anchor was invertible (that assumes
implicitly the existence of some action for the original
dynamics), the path integral can be explicitly taken in the bulk
of the topological  theory resulting in Feynman's partition
function for original action. With a general (non-invertible)
anchor, the answer for the partition function cannot be reduced to
the canonical form, but it remains fully consistent and allows
quite natural physical interpretation \cite{LS2}. If the anchor is
chosen to be zero, the partition function will correspond to the
classical transition amplitude \cite{Gozzietall}. The second
method to quantize a classical theory with non-Lagrangian
equations of motion \cite{LS2} suggests a nontrivial
generalization of the Schwinger-Dyson  equation that any partition
function must satisfy. This equation, involving classical
equations of motion and the Lagrange anchor, reduces  to the BV
quantum master equation whenever the anchor is invertible.

In this paper, we propose an alternative procedure of constructing
partition functions for general dynamical systems. This procedure
starts with the same input data: classical equations of motion and
the Lagrange anchor, but it exploits quite different idea and
technology. We call this procedure an \textit{augmentation}
because it is motivated by a widespread view that either a
non-Lagrangian system can be reshaped into an equivalent
Lagrangian model in an appropriately extended configuration space,
or it describes an effective dynamics emerging from a Lagrangian
theory after averaging over some degrees of freedom or their
exclusion from the equations of motion. So, the intuitive
intention about quantizing a non-Lagrangian theory is to augment
it first to a Lagrangian one, and then the augmented theory can be
quantized in the usual way. No general method is known to date to
\textit{equivalently} reformulate any given non-Lagrangian model
as a Lagrangian one by adding a finite number of new fields. We
propose a uniform procedure to construct an augmented Lagrangian
theory for any (non-)Lagrangian dynamics, which is not however an
equivalent reformulation. The augmented theory may have, in
principle, more degrees of freedom than the original model, but
classically, the original dynamics are easily singled out by
imposing appropriate boundary conditions on the extra fields.
These boundary conditions guarantee that the original fields
evolve precisely in the same way as in the original theory, while
new fields do not evolve at all. This reduction mechanism always
restores the original dynamics in the augmented theory including
the case where the original theory is Lagrangian. Quantizing the
augmented Lagrangian system by conventional BV procedure and
integrating the new fields out in the path integral, one gets the
original (not necessarily Lagrangian) dynamics quantized. If the
original theory is Lagrangian, the integral can be taken
explicitly over the augmentation fields with corresponding
boundary conditions, and the partition function obtained in this
way will coincide with that constructed from the BV master action
for the original Lagrangian. If the original theory is not
Lagrangian, the constructed partition function is still correct
that can be seen in several ways, although it cannot be
represented anymore as an exponential of any (local or non-local)
action functional.

Let us also comment on an essential  distinction between the
augmentation idea we use to quantize non-Lagrangian theories and
somewhat similar concept of ``auxiliary fields" \cite{{H1}}. The
fields are usually understood as auxiliary when they are
introduced to extend the dynamics in such a way that the extended
classical theory remains fully equivalent to the original one. In
particular this means that the number of independent initial data
for Cauchy problem remains the same as in the original theory. In
contrast to introducing the auxiliary fields, the augmentation
procedure results in a theory that has more degrees of freedom
than the original one. The extra dynamics are eliminated, however,
by imposing zero initial and/or boundary conditions on the
augmentation fields. At quantum level, these conditions provide
the absence of the ``augmenting particles'' in in- and out-states
of the quantum system. It might be relevant to mention that no
regular procedure is known yet for introducing auxiliary fields in
such a way as to convert any non-Lagrangian theory into an
equivalent Lagrangian one. In some specific models the way of
introducing auxiliary fields is known, although it often happens
that the restrictions are to be imposed strongly limiting the
admissible form of equations of motion\footnote{For example, in
higher-spin field theories, the auxiliary fields can be introduced
converting a non-Lagrangian model into Lagrangian one unless no
interaction has been switched on, even though the consistent
equations of motion with interaction are know for many years, see
\cite{Va} for a review and further references}. In contrast, the
augmentation is a regular procedure which always works well, given
equations of motion and à Lagrange anchor.

The paper is organized as follows.  To make the paper
self-contained, we review some recent developments in
path-integral quantization of non-Lagrangian theories that
includes basic definitions and some relevant statements from
\cite{KLS}, \cite{LS2}. In Section 2, we set up notation and
explain some basic facts concerning the general structure of not
necessarily Lagrangian gauge systems. We recall the notion of
Lagrange structure, which contains the Lagrange anchor as a key
ingredient, and put it in the context of $S_\infty$-algebras. The
corresponding Subsection 2.4 is addressed to the readership
familiar with basics of strongly homotopy algebras, others may
just omit this subsection. The paper can be further understood
without knowing the concept of $S_\infty$-algebras, although this
concept  provides a natural homological insight into the
quantization problem of non-Lagrangian dynamics. In Section 3, we
describe a BRST complex, which can be assigned to any
(non-)Lagrangian gauge system. As input data, this complex
involves the original equations of motion, generators of gauge
identities and gauge symmetries, and the Lagrange anchor. At
first, we define the ambient Poisson manifold that hosts this BRST
complex and construct the BRST charge by homological perturbation
theory. Further, we show that the BRST cohomology classes
precisely correspond to the physical observables of the original
(non-)Lagrangian theory. We also give an important interpretation
of this BRST complex as that resulting from the BFV-BRST
quantization of some constrained Hamiltonian system on the phase
space of original fields and their sources. Section 4 is devoted
to the quantization of the BRST complex. Quantizing the ambient
Poisson manifold, we define the quantum BRST cohomology and
present the generalized Schwinger-Dyson equation for the partition
function of a (non-)Lagrangian gauge theory. This equation is
shown to have a unique solution, which can be  written down in a
closed path-integral form.

Section 5 contains the main results of the paper. Namely, in
Section 5.1. we define the augmented BRST complex, which is build
on the original BRST complex and carries all the information about
the augmented theory. In Section 5.2. we unfold the structure of
the augmented BRST charge by interpreting it in terms of equations
of motion, gauge symmetry and Noether identity generators. As the
Lagrange anchor behind the augmented theory is always
nondegenerate (whatever the original anchor), the augmented
partition function has the standard Feynman's form. Moreover, the
corresponding action functional is proved to possess the property
of space-time locality provided the original equations of motion
and the Lagrange anchor do so. Finally, in Section 5.3. we present
an alternative path-integral representation for the quantum
averages of the original physical observables in terms of the
augmented action functional.

In Section 6, we apply the augmentation procedure to quantize two
non-Lagrangian field theories: Maxwell electrodynamics with
monopoles and self-dual p-forms. These models are known to admit
no Lagrangian formulation. However, we have found non-trivial
(degenerate) Lagrange anchors for these theories. Making use of
the anchors, we apply the augmentation method to construct
manifestly Poincar\'e invariant partition functions for both the
models.

\section{Lagrange structure and $S_\infty$-algebras}\label{rls}

\subsection{Classical dynamics}

In field theory one usually  deals with the space $Y^X$ of all
smooth maps from a space-time manifold $X$ to a target manifold
$Y$. The atlases of coordinate charts  on $X$ and $Y$ define then
a natural atlas on $Y^X$ such that each map  $x: X\rightarrow Y$
is specified locally by a set of smooth fields $x^i$, the
coordinates on the infinite-dimensional manifold $Y^X$. Hereafter
we use De Witt's condensed notation \cite{DW}, whereby the
superindex ``$i$'' comprises both the local coordinates on $X$ and
the discrete indices labelling the components of the field $x$. As
usual, the superindex repeated implies summation over the discrete
indices and integration over the space-time coordinates w.r.t. an
appropriate measure on $X$. The partial derivatives
$\partial_i=\partial/\partial x^i$ are understood as variational
ones.

In the context of local field theory, the space  $Y^X$ is known as
the \textit{space of all histories} and the \textit{true
histories} are specified by a set of PDE's
\begin{equation}\label{T}
    T_a(x)=0\,.
\end{equation}
Here we do not assume the field equations  to come from the least
action principle, hence the indices $i$ and $a$, labelling the
fields and equations, may run through completely different sets.
In the case where $X$ is a manifold with boundary, Eqs. (\ref{T})
are also supplemented with a suitable set of boundary conditions.
Usually, the boundary conditions specify the values of fields
and/or their derivatives up to some fixed order.  Varying these
values, collectively called the \textit{boundary data},  one gets
a family of different solutions to Eqs. (\ref{T}).

For our purposes it is convenient to think of $T=\{T_a(x)\}$ as a
section of some vector bundle $\mathcal{E}\rightarrow M$ over
subspace of all fields  $M\subset Y^X$ with given boundary data.
Then the set of all true histories $ \Sigma$  belonging to $M$ is
identified with zero locus of $T\in \Gamma(\mathcal{E})$:
\begin{equation}\label{}
    \Sigma=\{x\in M| \; T(x)=0\;\}\,.
\end{equation}
Using the physical terminology, we refer to $\Sigma$ as the
\textit{shell}. Under the standard regularity conditions
\cite{HT}, $\Sigma\subset M$ is a smooth submanifold associated
with an orbit of gauge symmetry transformations (see Eq.(\ref{RT})
below); in the absence of gauge symmetries  the shell $\Sigma$
is  a single point of $M$. In the following  we will
always assume $\Sigma$ to be a \textit{connected} submanifold for
each choice of boundary data.

Thus, the classical dynamics are completely specified by a section
$T$ of some vector bundle $\mathcal{E}\rightarrow M$ over the
space of all histories subject to boundary conditions. For this
reason we call $\mathcal{E}$ the \textit{dynamics bundle}.

\subsection{Regularity conditions} To avoid pathological examples, some regularity
conditions are usually imposed on a classical system.  To
formulate these conditions in an explicitly covariant way, let us
introduce an arbitrary connection $\nabla$ on $\mathcal{E}$ and
define the section
\begin{equation}
J=\nabla T \in \Gamma(T^\ast M\otimes \mathcal{E})
\end{equation}
This section, in turn, defines the $M$-bundle morphism
\footnote{To simplify notation, we will not  distinguish between
an $M$-bundle morphism $H: \mathcal{E}\rightarrow \mathcal{E}'$,
the induced homomorphism $\Gamma(H):
\Gamma(\mathcal{E})\rightarrow \Gamma(\mathcal{E}')$ on sections,
and the associated section $\widetilde{H}\in
\Gamma(\mathcal{E}^\ast\otimes\mathcal{E}')\simeq
\mathrm{Mor}(\mathcal{E}, \mathcal{E}')$, denoting all these maps
by one and the same letter $H$.}
\begin{equation}\label{}
J: TM\rightarrow \mathcal{E}\,,
\end{equation}
which is not necessarily of constant rank.

\begin{defn}\label{Reg}
A classical system $(\mathcal{E},T)$ is said to be \textit{regular
of type} $(m,n)$, if there exists a finite sequence of vector
bundles $\mathcal{E}_k\rightarrow M$ and $M$-bundle morphisms
\begin{equation}\label{exseq}
   0\rightarrow \mathcal{E}_{-m}{\rightarrow}\cdots\rightarrow
   \mathcal{E}_{-1}\stackrel{R}{\longrightarrow}TM\stackrel{J}{\longrightarrow}\mathcal{E}\stackrel{Z}
   {\longrightarrow}\mathcal{E}_{1}{\rightarrow}
    \cdots{\rightarrow}\mathcal{E}_{n}\rightarrow 0
\end{equation}
satisfying conditions:
\begin{enumerate}
    \item [(a)]there is a tubular  neighbourhood $U\subset M$ of $\Sigma$ such
    that all the morphisms (\ref{exseq}) have constant ranks over $U$;
    \item [(b)]upon restriction to $\Sigma$, the chain (\ref{exseq}) makes an exact sequence.
\end{enumerate}
\end{defn}

This definition has several important corollaries elucidating its
meaning:

\begin{corollary}\label{RS}
The shell $\Sigma\subset M$ is a smooth submanifold with
${T\Sigma}=\mathrm{Im}\,R|_\Sigma$.
\end{corollary}

\begin{corollary}\label{vanishing}
For any vector bundle $\mathcal{V}\rightarrow M$ and a section
$K\in \Gamma(\mathcal{V})$ vanishing on $\Sigma\subset M$, there
is a smooth section $W\in \Gamma(\mathcal{E}^\ast\otimes
\mathcal{\mathcal{V}}) $ such that
\begin{equation}
K=\langle T, W\rangle\,,
\end{equation}
where the triangle brackets denote contraction of $T$ and $W$.
Informally speaking, any  on-shell vanishing section is
proportional to $T$.
\end{corollary}

\begin{corollary}\label{Z0} When exist, the morphisms (\ref{exseq}) are not unique
off shell. Thinking of these morphisms as the sections of the
corresponding  vector bundles, one can add to them any sections
vanishing on $\Sigma$, leaving the properties (a),(b) unaffected.
In particular, by making a shift
\begin{equation}
Z\rightarrow Z+Z_0\,,\qquad Z_0|_{\Sigma}=0\,,
\end{equation}
if necessary, we can always assume that $T \in \ker Z$. In view of
the previous remark, the section $Z_0$ is proportional to $T$.
\end{corollary}

\begin{corollary}In the definition above we can pass from the sequence
(\ref{exseq}) to the transpose one by replacing each vector bundle
with its dual and inverting all the arrows. The transpose sequence
will  meet the same conditions (a),(b) as the original one.
\end{corollary}

In this paper we deal mostly with the quantization of regular
$(1,1)$-type Lagrange structures associated to the four-term
sequences
\begin{equation}\label{4t-seq}
    0\rightarrow \mathcal{F}\stackrel{R}{\longrightarrow}TM
    \stackrel{J}{\longrightarrow}\mathcal{E}\stackrel{Z}
    {\longrightarrow}\mathcal{G}\rightarrow
    0\, .
\end{equation}
The on-shell exactness at $TM$ suggests that for any vector field
$V\in \Gamma(TM)$ obeying condition $\nabla_V T|_\Sigma=0$ there
exists a section $\varepsilon\in \Gamma(\mathcal{F})$ such that $V
=R(\varepsilon)$. Combining this with Corollary \ref{vanishing},
we can write
\begin{equation}\label{RT}
    R^i_\alpha\nabla_iT_a=U^b_{\alpha a}T_b
\end{equation}
for some $U\in \Gamma(\mathcal{E}\otimes
\mathcal{E}^\ast\otimes\mathcal{F})$. Here indices $a$, $i$,
$\alpha$ label the components of the corresponding sections w.r.t.
to some frames $\{e^\alpha\} \in \Gamma(\mathcal{F}|_{U})$,
$\{e^a\}\in \Gamma(\mathcal{E}|_{U})$, and  $\{\partial_i\} \in
\Gamma(TU)$ associated with a trivializing coordinate chart
$U\subset M$. Let $\{e^A\}$ be a frame in $\mathcal{G}$ over $U$.
In view of Corollary \ref{Z0} the on-shell exactness at term
$\mathcal{E}$ implies then
\begin{equation}\label{ZT}
    Z_A^aT_a=0\,,
\end{equation}
if  $Z$ was chosen in an appropriate way.   Relations  (\ref{RT})
and (\ref{ZT}) have a straightforward interpretation in terms of
constrained dynamics \cite{HT}:  the homomorphism $R$ is
identified with an irreducible set of gauge symmetry
generators for the classical equations of motion $T=0$, while the
homomorphism $Z$ generates a set of independent Noether's
identities. Having in mind this interpretation, we term
$\mathcal{F}$ and $\mathcal{G}$ the \textit{gauge algebra bundle}
and the \textit{Noether identity bundle}, respectively.  Notice
that the irreducibility of the gauge symmetry generators is
provided by the on-shell exactness of (\ref{4t-seq}) at
$\mathcal{F}$, while irreducibility of Noether's identity
generators follows from the on-shell exactness of the transpose of
(\ref{4t-seq}) at $\mathcal{G}^\ast$.

This interpretation of homomorphisms $R$ and $Z$ applies to the
general regular systems of type $(m,n)$, except that the bases of
the gauge algebra and Noether's identity generators may be
overcomplete  (reducible). A general $(n+1,m+1)$-type gauge
theory with $n>0$ and/or $m>0$ corresponds to the case of
$n$-times reducible generators of gauge transformations and/or
$m$-times reducible generators of Noether's
identities. The theories of type $(0,0)$ are described  by
linearly  independent equations of motion having a unique
solution.

\subsection{Lagrange structure} In the context of covariant path-integral quantization, the passage from  classical to
quantum theory involves, besides classical equations of motion,
one more geometric ingredient called the Lagrange structure
\cite{KLS}.

\begin{defn}\label{LS} Given a classical system $(\mathcal{E},T)$, a \textit{{Lagrange structure}} is an
$\mathbb{R}$-linear map $d_\mathcal{E}: \Gamma(\wedge^n
\mathcal{E})\rightarrow \Gamma(\wedge^{n+1}\mathcal{E})$ obeying
two conditions:
\begin{enumerate}
      \item [(i)] \,  $d_\mathcal{E}$ is a derivation of degree
      1, i.e.,
$$
 d_{\mathcal{E}}(A\wedge B)=d_\mathcal{E}A\wedge B +
(-1)^{n}A\wedge d_\mathcal{E} B\,,
$$
for any $A\in \Gamma(\wedge^n \mathcal{E})$ and $ B\in
\Gamma(\wedge^\bullet \mathcal{E})$;
 \item [(ii)] \,
$d_\mathcal{E} T=0$\,.
\end{enumerate}
Here we identify $\Gamma(\wedge^0\mathcal{E})$ with $C^\infty(M)$.
\end{defn}

\vspace{5mm} Due to the Leibnitz rule (i), in each
trivializing chart $U\subset M$ the operator $d_\mathcal{E}$ is
completely specified by its action on the coordinate functions $x^i$
and the basis sections $e^a$ of $\mathcal{E}|_U$:
\begin{equation}\label{dE}
    d_\mathcal{E} x^i=V^i_a(x) e^a\,,\qquad d_\mathcal{E} e^a=\frac12C_{bc}^a(x)e^b\wedge
    e^c\,.
\end{equation}
Applying $d_\mathcal{E}$ to the section $T=T_ae^a$, one can see
that the property (ii) is equivalent to the following  structure
relations:
\begin{equation}\label{Brel}
d_\mathcal{E} T =(V^i_a\partial_i T_b - C_{ab}^cT_c )e^a\wedge
e^b=0\,.
\end{equation}
The first relation in (\ref{dE}) means also that $d_\mathcal{E}$
defines a bundle homomorphism $V: \mathcal{E}^\ast\rightarrow TM$.
The section  $V\in \Gamma (\mathcal{E}\otimes TM)$ is called the
\textit{Lagrange anchor}.

\begin{defn}
A Lagrange structure $(\mathcal{E},T,d_\mathcal{E})$ is said to be
\textit{regular} at $p\in M$, if there exists a vicinity $U\subset
M$ of  $p$ such that the $M$-bundle morphism
\begin{equation}\label{comple}
R\oplus V: \mathcal{E}_{-1}\oplus \mathcal{E}^\ast \rightarrow TM
\end{equation}
has a constant rank\footnote{Of course, in the context of
infinite-dimensional manifolds the notion of rank needs
clarification. An appropriate definition can be done, for example,
in the case of local field theories.} $r$ over $U$.   The number
$r$ is called the \textit{rank of Lagrange structure at} $p\in M$.
The Lagrange structure is said to be \textit{complete} at $p$, if
the homomorphism (\ref{comple}) is surjective  on $U$. Finally, we
say that the Lagrange structure is regular (or complete), if it is
regular (or complete) at any point of $M$.
\end{defn}

\begin{rem}
In view of Definition \ref{Reg}, the regularity of the Lagrange
structure at  $p\in \Sigma$ is equivalent to the regularity at $p$
of the anchor morphism $V: \mathcal{E}^\ast\rightarrow TM$, i.e.,
there exists  a sufficiently small vicinity $U\subset M$ of $p\in
\Sigma$ such that $V$ has constant rank over $U$.
\end{rem}

\begin{rem}
In the context of quasiclassical quantization we will deal with in
sequel, it is also appropriate to introduce the notions of
\textit{weakly regular} and \textit{weakly complete Lagrange
structures} by requiring regularity and completeness only for the
points of  $\Sigma$.
\end{rem}

\begin{thm}[ Splitting theorem \cite{KLS}]\label{LA}
Let $p\in M$ be a regular point of the Lagrange structure
$(\mathcal{E},T,d_\mathcal{E})$, then there is a coordinate system
$(y^1,...,y^r, z^1,...,z^k)$ centered at $p$ together with a set
of local functions $S(y)$, $E^1(y),...,E^k(y)$ such that equations
$T_a(y,z)=0$ are equivalent to
$$
\frac{\partial S(y)}{\partial y^{I}}=0\,,\qquad
    z^J = E^J(y)\,,
$$
and the Lagrange anchor $V=(V^J, V_I)$ is given by the abelian
vector distribution
$$
  V^J=0\,,\qquad V_I=\frac{\partial}{\partial y^I}+\frac{\partial E^J}{\partial
  y^I}\frac{\partial}{\partial z^J}\,.
$$
Here the number $r$ is the rank of the Lagrange structure at $p\in
M$.
\end{thm}

In case $r<\dim M$, it is  natural to call $S(y)$ a
\textit{partial action}.

Although the theorem above  ensures the split of local coordinates
into ``Lagrangian'' $y$'s and ``non-Lagrangian'' $z$'s, it is by
no means necessary to explicitly perform this splitting in order
to develop the theory further. The subsequent formulas do not
involve such a split. Moreover, the method is insensitive to the
rank of the Lagrange anchor producing  a well-defined
path-integral quantization in the irregular case as well.

\vspace{3mm}
\begin{example} Let us illustrate the definitions above
by an example of a Lagrangian gauge theory with action $S(x)$. The
equations of motion read
\begin{equation}\label{dS}
   T\equiv dS(x)=0\,,
\end{equation}
so that the dynamics bundle $\mathcal{E}$ is given by the
cotangent bundle $T^\ast M$ of the space of all histories. The
canonical Lagrange structure, resulting in standard quantization,
is given by the exterior  differential $d: \Gamma(\wedge^n T^\ast
M)\rightarrow \Gamma(\wedge^{n+1}T^\ast M)$. The defining
condition for the Lagrange structure (\ref{Brel}) takes the form
\begin{equation}\label{}
    dT=d\,{}^2S\equiv 0.
\end{equation}
The Lagrange anchor is defined by the identical homomorphism
\begin{equation}\label{V1}
V=\mathrm{id}: TM\rightarrow TM\,,
\end{equation}
and hence the Lagrange structure is regular and complete. Suppose
the action $S$ is gauge invariant. Then there exist a set of gauge
algebra generators defining an $M$-bundle morphism $R:
\mathcal{F}\rightarrow TM$ such that
\begin{equation}\label{RdS}
    \langle R(\varepsilon),dS\rangle=0
\end{equation}
for any gauge parameter $\varepsilon\in\Gamma( \mathcal{F})$. So,
equations  (\ref{dS}) appear to be linearly dependent.
Differentiating the last identity w.r.t. some connection $\nabla$
on $\mathcal{F}\otimes TM$,  we arrive at Rel. (\ref{RT}) with
$U_{\alpha i}^j =\nabla_i R_\alpha^j$.

Thus, we see that for ordinary Lagrangian gauge theories the
dynamics bundle coincides with the cotangent bundle
($\mathcal{E}=T^\ast M$), the Noether identity bundle coincides
with the gauge algebra bundle ($\mathcal{F}=\mathcal{G}$), and the
generators of gauge symmetry coincide with the generators of
Noether's identities ($R=Z$). For a general regular system of
type $(1,1)$ neither of these coincidences should necessarily
occur. For instance, it is possible to have gauge  invariant, but
linearly independent equations of motion; and conversely, a theory
may have linearly dependent equations of motion without gauge
symmetry.

\end{example}

\subsection{$S_\infty$-algebras} Recall that the conventional BV
formalism for a Lagrangian gauge theory starts by introducing
ghost fields to every gauge symmetry, and then an antifield for
every field. The space of all fields and antifields is endowed
with the canonical odd Poisson bracket $( \, \cdot \, , \cdot \,
)$ and the original action functional is extended to the master
action $S$ defined as a proper  solution to the classical master
equation $(S,S)=0$. The classical BRST differential $Q =(S, \,
\cdot )$, being a nilpotent derivation of the odd Poisson algebra
of functions, incorporates then  both the dynamical equations and
the gauge algebra structure. Thus, the odd Poisson geometry
provide a natural framework for the BV field-antifield
formalism\footnote{In the physical literature the odd Poisson
manifolds are usually called \textit{anti-Poisson} manifolds.
Correspondingly, the odd Poisson brackets are referred to as
\textit{antibrackets}. On the other hand, in mathematics the odd
Poisson algebras (brackets) are also known under the names  of
Schouten or Gerstenhaber algebras (brackets).}. In \cite{KLS}, we
have shown that the quantization of general non-Lagrangian gauge
theories call for a strongly homotopical version of the odd
Poisson algebras.

\begin{defn}
An $S_\infty$-algebra ($S$ for Schouten) is a
$\mathbb{Z}_2$-graded, supercommutative, and associative algebra
$A$ endowed with a sequence of odd linear maps $S_n: A^{\otimes
n}\rightarrow A$ such that
\begin{enumerate}
    \item[(a)]
    $S_n(...,a_k,a_{k+1},...)=(-1)^{\epsilon(a_k)\epsilon(a_{k+1})}S_n(...,a_{k+1},a_k,...)$,\\
    $\epsilon(a)$ being the parity of a homogeneous element $a\in
    A$.
    \item [(b)]$a \mapsto S_n(a_1,...,a_{n-1},a)$ is a derivation
    of $A$ of the parity\\ $1+\sum_{k=1}^{n-1}\epsilon(a_k)\; (\mathrm{mod}\;
    2)$.
    \item [(c)]For all $n\geq 0$,
$$
\sum_{k+l=n}\sum_{(k,l)-\mathrm{shufle}} (-1)^\epsilon S_{l+1}
(S_k(a_{\sigma
(1)},...,a_{\sigma(k)}),a_{\sigma(k+1)},...,a_{\sigma(k+l)})=0\,,
$$
where $(-1)^\epsilon $ is the natural sign prescribed by the sign
rule for permutation of homogeneous elements $a_1,...,a_n\in A$.
\end{enumerate}
Recall that a $(k,l)$-shuffle is a permutation of indices
$1,2,...,k+l$ satisfying $\sigma(1)<\cdots < \sigma(k)$ and
$\sigma(k+1)<\cdots<\sigma(k+l)$.
\end{defn}

When $S_0=0$ we speak  about a \textit{flat} $S_\infty$-algebra.
In that case $S_1: A\rightarrow A$ is a differential with
$(S_1)^2=0$, and $S_2$ induces an odd Poisson structure on the
cohomology of $S_1$. An odd Poisson algebra can be
regarded as an $S_\infty$-algebra with bracket $S_2:A\otimes
A\rightarrow A$ and all other $S_k=0$. In fact, properties (a) and (c)
characterize $L_\infty$-algebras. We refer to \cite{Vo} for a
recent discussion of $S_\infty$-algebras.

It turns out that any  Lagrange structure of type $(m,n)$ gives
rise to a flat $S_\infty$-algebra on the supercommutative algebra
of sections

\begin{equation}\label{A}
A= \Gamma\big(\wedge^\bullet\mathcal{E}\otimes\bigotimes_{k=1}^m
S^\bullet (\Pi^k \mathcal{E}_{-k})\otimes\bigotimes_{l=1}^n
    S^\bullet(\Pi^{l+1}
    \mathcal{E}_l)\big)\,.
\end{equation}
Here $S^\bullet $ stands for symmetric tensor powers (in the
$\mathbb{Z}_2$-graded sense) and $\Pi$ denotes the parity
reversion operation, i.e., $\Pi \mathcal{E}$ is a vector bundle
over $M$ whose fibers are odd linear spaces. By definition,
$\Pi^2=\mathrm{id}$ and $S^\bullet(\Pi \mathcal{E})=\wedge^\bullet
\mathcal{E}$.

In the next section, applying the machinery of BRST theory, we
give an explicit description for $S_\infty$-algebras associated
with $(1,1)$-type Lagrange structures. Extension to the general
Lagrange structures is straightforward.

\section{BRST complex}

\subsection{An ambient symplectic supermanifold} Let $(\mathcal{E},T,d_\mathcal{E})$
 be a regular Lagrange structure corresponding to the four-term
sequence (\ref{4t-seq}). Following the general line of ideas of
BRST theory, we realize $M$ - the space of all histories - as the
body of a graded supermanifold $\mathcal{N}$. The latter is chosen to be the total space
of the following  graded vector bundle over $M$:
\begin{equation}\label{N}
\Pi (\mathcal{F}\oplus \mathcal{F}^\ast)\oplus T^\ast M \oplus
\Pi(\mathcal{E}\oplus \mathcal{E}^\ast)\oplus (\mathcal{G}\oplus
\mathcal{G}^\ast) \, .
\end{equation}
Here $\mathcal{F}$, $\mathcal{E}$, and $\mathcal{G}$ are the
bundles of gauge algebra, dynamical equations and the Noether
identities, respectively. The base $M$ is imbedded into (\ref{N})
as the zero section. In addition to the Grassman parity the fibers
of (\ref{N}) are graded by \textit{ghost number} valued in
integers. To avoid cumbersome sign factors, we will assume the
base $M$ to be an ordinary (even) manifold that corresponds to the
case of gauge systems without fermionic degrees of freedom. Then
the Grassman parities of fibers correlate with their ghost numbers
in a rather simple way: the even coordinates have even ghost
numbers, while the odd coordinates have odd ghost numbers. The
supermanifold $\mathcal{N}$ is also endowed with an
$\mathbb{N}$-grading called the \textit{momentum degree} (or
$m$-degree for short).

It is convenient to arrange the information about all the
aforementioned gradings of local coordinates in a single  table:

$$
\begin{tabular}{|l|c|c|c|c|c|c|c|c|}
  \hline

base and fibers   &$M$ & $T^\ast M$ & $\mathcal{F}
$&$\mathcal{F}^\ast$ &$\mathcal{E}$&$
\mathcal{E}^\ast$&$\mathcal{G}$& $\mathcal{G}^\ast$\\
\hline

local coordinates           &\,$x^i$ & $\bar x_j$&\, $c^\alpha
$\,& \,$\bar c_{\beta}$\, &\, $\eta_a$ & \,$\bar
\eta^b$&\,$\xi_A$\,&\,$\bar
           \xi^B$\\
  \hline
  $\epsilon$=Grassman parity            & 0 & 0 & 1 & 1 & 1 & 1 & 0 & 0\\
  \hline
  $\gh$ = ghost number                   & 0 & 0 & 1 & -1 & -1 & 1 & -2&2\\
  \hline
  $\mathrm{Deg}$ = momentum degree      & 0 & 1 & 0 & 1 & 0 & 1&0&1 \\
  \hline
\end{tabular}
$$
\begin{center}
{Table 1}
\end{center}
Upon splitting all the coordinates into the ``position
coordinates'' $\varphi^I=(x^i,c^\alpha, \eta_a, \xi_A)$ and
``momenta'' $\bar \varphi_J=(\bar x_i, \bar c_\alpha, \bar \eta^a,
\bar\xi^A)$ the assignment of gradings becomes easy to see
\begin{equation}\label{}
\begin{array}{ll}
    \gh(\bar\varphi_I)=-\gh(\varphi^I)\,,&\qquad
    \epsilon(\bar\varphi_I)=\epsilon(\varphi^I)\,,\\[3mm]
    \mathrm{Deg}
    (\bar\varphi_I)=1\,,&\qquad \mathrm{Deg}(\varphi^I)=0\,.
    \end{array}
\end{equation}

Let us denote by $C^{\infty}(\mathcal{N})$ the supercommutative
algebra of ``smooth functions'' on $\mathcal{N}$. By definition,
the generic element of $C^\infty(\mathcal{N})$ is given by a
formal power series in the fiber coordinates with coefficients in
$C^\infty(M)$.

Fixing a linear connection
$\nabla=\nabla_{\mathcal{F}}\oplus\nabla_\mathcal{E}\oplus
\nabla_\mathcal{G}$ on $\mathcal{F}\oplus \mathcal{E}\oplus
\mathcal{G}$,  we endow $\mathcal{N}$ with the exact symplectic
structure
\begin{equation}\label{Theta}
  \omega=d(\bar x_idx^i + \bar
    c_{\alpha}\nabla c^\alpha +\bar\eta^a \nabla \eta_a + \bar\xi^A\nabla
    \xi_A)\,,
    \end{equation}
    where
    \begin{equation}
\nabla c^\alpha=dc^\alpha+dx^i\Gamma_{i \beta}^\alpha c^\beta\,,
\end{equation}
and similar expressions are assumed for covariant differentials of
$\eta$'s and $\xi$'s. Thus,  $C^\infty(\mathcal{N})$ becomes a
Poisson algebra; the nonvanishing  Poisson brackets of local
coordinates are given by
\begin{equation}\label{brcov}
\begin{array}{llll}
 \{\bar\eta^b,\eta_a\}=\delta_a^b \,,\;&  \{\bar x_i, \eta_a\}=\Gamma_{i a}^b
 \eta_b
 \,,&&\hspace{-4mm}
 \{\bar x_i,\bar\eta^b\}=-\Gamma_{ia}^b\bar\eta^a\,, \\[3mm]
 \{\bar c_\alpha,c^\beta\}=\delta^\beta_\alpha\,,\;& \{\bar x_i,c^\alpha \}=\Gamma_{i\beta}^\alpha c^\beta\,
 \,,&&\hspace{-4mm}
 \{\bar x_i,\bar c_\beta\}= -\Gamma_{i\beta}^\alpha \bar c_\alpha\,,
  \\[3mm]
  \{\bar\xi^A,\xi_B\}=\delta_B^A\,,\;& \{\bar
  x_i,\xi_A\}=\Gamma_{iA}^B\xi_B\,,\quad&&\hspace{-4mm}
  \{\bar
  x_i,\bar \xi^A\}=-\Gamma_{iB}^A\bar\xi^B\,,\\[3mm]
\{\bar x_i, x^j\}=\delta_i^j\,, \;& \{\bar x_i,\bar
x_j\}=R_{ija}^b \bar\eta^a\eta_b&\hspace{-4mm}+&\hspace{-4mm}
  R_{ij\alpha}^\beta c^\alpha\bar
  c_\beta+R_{ijA}^B\bar\xi^A\xi_B\,.
  \end{array}
\end{equation}
Here the structure functions determining the Poisson brackets of $\bar
x_i$ and $\bar x_j$ are just the components of the curvature
tensor of $\nabla$.

Notice that the equations $\bar \varphi_I=0$ define the Lagrangian
submanifold
\begin{equation}\label{}
\mathcal{L}=\Pi (\mathcal{F}\oplus  \mathcal{E})\oplus \mathcal{G}
\subset \mathcal{N}\,,
\end{equation}
and the supercommutative algebra of functions
$C^\infty(\mathcal{L})$ is naturally isomorphic to the algebra
(\ref{A}) with $m=n=1$.

\subsection{BRST charge}
It turns out that all the ingredients of a classical gauge system
as well as a Lagrange structure can be naturally incorporated into
a single object $\Omega$, the classical BRST
charge\footnote{Relevance of this terminology is explained in the
next subsection.}. By definition \cite{KLS}, the BRST charge
$\Omega$ is an element of the Poisson algebra
$C^\infty(\mathcal{N})$ such that
    \begin{itemize}
        \item [(i)] $\epsilon(\Omega)=1$\,,
        $\quad\mathrm{gh}(\Omega)=1$,
        $\quad\mathrm{Deg}(\Omega)>0$;\vspace{2mm}
        \item  [(ii)] $\Omega = \bar\eta^aT_a+c^\alpha R_\alpha^i\bar x_i+\bar\xi^AZ_A^a\eta_a+
        \bar\eta^aV_a^i\bar x_i+\cdots$;\vspace{2mm}
        \item [(iii)] $\{\Omega,\Omega\}=0$\,.
    \end{itemize}
The dots in (ii) refer to the terms which are at least linear in
$\eta_a$ and $\bar c_\alpha$ or at least quadratic in $\bar x_i$.
Equation (iii) is known as the (classical)  master equation.
Conditions (i)-(iii) determine $\Omega$ up to a canonical
transformation of $C^\infty(\mathcal{N})$. The existence of
$\Omega$ is proved by standard tools of homological perturbation
theory \cite{KLS}.

It is instructive to consider expansion of $\Omega$ in powers
of momenta. In view of (i) the expansion starts with terms linear
in $\bar\varphi$, i.e.,
\begin{equation}\label{exp}
    \Omega = \sum_{k=1}^\infty \Omega_k\,,\qquad \mathrm{Deg}(\Omega_k)=k\,.
\end{equation}
On substituting (\ref{exp}) into the master equation (iii), we get
\begin{equation}\label{me}
    \{\Omega_1,\Omega_1\}=0\,,\quad
    \{\Omega_1,\Omega_2\}=0\,, \quad \{\Omega_2, \Omega_2\}=-2\{\Omega_2,\Omega_3\}\,,\quad \mathrm{etc}\,.
\end{equation}
We see that the leading term $\Omega_1=\Omega^I\bar\varphi_I$
gives rise to the \textit{homological vector field} on
$\mathcal{L}$,
\begin{equation}\label{Q}
    Q\equiv \Omega^I\frac{\partial}{\partial \varphi^I} =
    T_a\frac{\partial}{\partial \eta_a}+c^\alpha
    R_\alpha^i\frac{\partial}{\partial x^i}+\eta_aZ^a_A\frac{\partial}{\partial
    \xi_A}+\cdots\,,
\end{equation}
which carries all the information about the classical system
itself, with no regard to the Lagrange structure\footnote{In the
usual BV theory the operator $Q$ is known as the classical BRST
differential \cite[\S 8.5]{HT}.}. Evaluating the nilpotency
condition $Q^2=0$ to  lowest order in fiber coordinates, one
immediately recovers Rels.(\ref{RT}, \ref{ZT}) characterizing
$T=0$ as a set of gauge invariant and linearly dependent equations
of motion, with $R$ and $Z$ being the generators of gauge
transformations and Noether identities, respectively.

The Lagrange anchor $V: \mathcal{E}^\ast\rightarrow TM$ defining
the Lagrange structure for the classical system (\ref{Q}) enters
the next term
\begin{equation}
\Omega_2=\Omega^{IJ}(\varphi)\bar\varphi_I\bar\varphi_J=\bar\eta^aV_a^i\bar
x_i+\cdots\,.
\end{equation}
Relations (\ref{me}) characterize $\Omega_2$ as a weak
anti-Poisson structure on $\mathcal{L}$, i.e., $Q$-invariant, odd
bivector field satisfying the Jacobi identity up to homotopy. The
corresponding ``weak''  antibracket reads
\begin{equation}\label{wabr}
(a,b)\equiv\{\{\Omega_2,a\},b\}\,,\qquad a,b \in
C^{\infty}(\mathcal{L})\,.
\end{equation}
Examining  the Jacobi identity for this bracket, one finds
\begin{equation}\label{}
\begin{array}{c}
    (a,(b,c))+(-1)^{\epsilon(b)\epsilon(c)}((a,c),b)+
    (-1)^{\epsilon(a)(\epsilon(b)+\epsilon(c))}((b,c),a)=\\[3mm]
- S_3( Qa,b,c)-(-1)^{\epsilon(a)\epsilon(b)}S_3(a,Qb,c)
-(-1)^{(\epsilon(a)+\epsilon(b))\epsilon(c)}S_3(a,b,Qc)\\[3mm]-
QS_3(a,b,c)\,,
    \end{array}
\end{equation}
where we have introduced the following notation:
\begin{equation}\label{Sn}
    S_n(a_1,a_2,...,a_n)\equiv\{...\{\{\Omega_n,a_1\}a_2\},...,a_n\}\,,\qquad a_k\in
    C^{\infty}(\mathcal{L})\,.
\end{equation}
Evidently, the weak antibracket (\ref{wabr}) induces a genuine
antibracket in the $Q$-cohomology.

It is Rel.  (\ref{Sn}) that  defines the aforementioned
$S_\infty$-structure on the supercommutative algebra
$C^\infty(\mathcal{L})$: By definition, each $S_n$ is a symmetric
multi-differentiation of $C^\infty(\mathcal{L})$ and the
generalized Jacobi identities for the collection of maps $\{S_n\}$
readily follow  from the master equation $\{\Omega,\Omega\}=0$ for
the BRST charge. Since $\mathrm{Deg}\,(\Omega)>0$, this
$S_\infty$-algebra is flat.

\subsection{Hamiltonian interpretation} In the conventional BFV
approach, the BRST charge arises as a tool for quantizing
first-class constrained Hamiltonian systems. A glance at  Table 1
is enough to see that the spectrum of ghost numbers corresponds to
that of the BFV-BRST formalism for a first-class constrained
Hamiltonian system with linearly dependent constraints \cite{HT}.
In order to make this interpretation more explicit, let us combine
the local coordinates with ghost numbers $1$ and $-1$ into the
ghost coordinates $\mathcal{C}^I=(\bar \eta^a, c^\alpha)$ and
ghost momenta $\bar{\mathcal{P}}_I=(\eta_a, \bar c_\alpha)$,
respectively. In this notation the above BRST charge (\ref{exp})
can be rewritten as
\begin{equation}\label{Otheta}
    \Omega = \mathcal{C}^I\Theta_I(x,\bar x)+ \bar{\mathcal{P}}_I\Xi_A^I(x,\bar
    x)\xi^A+ \frac12 \bar{\mathcal{P}}_K {U}^K_{IJ}(x,\bar
    x)\mathcal{C}^J\mathcal{C}^I + o(\bar{\mathcal{P}}^2,\xi^2)\,,
\end{equation}
where the expansion coefficients  $\Theta_I
=({\widetilde{T}}_a,\widetilde{R}_\alpha)$ and
$\Xi^I_A=(\widetilde{Z}_A^a,Z_A^\alpha)$, playing the role of
first-class constraints and their null-vectors, are given by the
formal power series in $\bar x$'s:
\begin{equation}\label{TRZ}
\begin{array}{l}
\widetilde{T}_a(x,\bar x)=T_a(x)+V_a^i(x)\bar x_i+o(\bar x^2)\,,\\[3mm]
\widetilde{R}_\alpha(x,\bar x)=R_\alpha^i(x)\bar x_i+o(\bar
x^2)\,,\\[3mm]
\widetilde{Z}_A^a(x,\bar x)=Z_A^a(x)+o(\bar x)\,.
\end{array}
\end{equation}
To lowest order in $\mathcal{C}$'s,  Eqs. (\ref{me}) reproduce
the standard involution relations for a set of reducible
first-class constraints w.r.t. the canonical Poisson bracket on $T^\ast M$:
\begin{equation}\label{ThetaInv}
    \{\Theta_I,\Theta_J\}={U}_{IJ}^K\Theta_K\,,\qquad
    \Xi_A^I\Theta_I=0\,.
\end{equation}
{}From the regularity condition it follows immediately that the
number of independent  first-class constraints $\Theta_I \approx
0$ is equal to $\dim M$. In physical terms, one can interpret this
fact concluding that the Hamiltonian system under consideration
has no physical degrees of freedom. From the geometrical
viewpoint, this implies  that the equations $\Theta_I=0$ define a
Lagrangian submanifold $L\subset T^\ast M$; more accurately, $L$
is a formal Lagrangian submanifold as we are not concerned with
convergence of the formal series (\ref{TRZ}).

One can also regard the constraints $\Theta_I \approx 0$ as a
formal deformation of those given by the leading terms of
expansions (\ref{TRZ}) in the ``direction'' of the Lagrange anchor
$V$. From this standpoint, the Lagrange structure is just the
infinitesimal of deformation of the Lagrangian submanifold $L_0
\subset T^\ast M$ defined by the ``bare'' first-class constraints
$T_a(x) \approx 0$ and $ R^i_\alpha(x)\bar x_i \approx0$.

Associated with the first-class constraints $\Theta_I\approx 0$ is
the Hamiltonian action on the cotangent bundle of the space of all
histories
\begin{equation}\label{top}
    S[\lambda, x,\bar x]=\int_{t_1}^{t_2}dt(\bar x_i\dot{x}^i
    -\lambda^I\Theta_I(x,\bar x))\,.
\end{equation}
The action describes a pure topological field theory having no
physical evolution w.r.t. to $t$. It should be emphasized, that the
``time" $t$  is an auxiliary $(d+1)$-st dimension, which has
nothing to do  with the evolution parameter in the (differential)
equations of motion $T_a=0$. The true physical time is among the
original $d$ dimensions.

The model (\ref{top}) is invariant under the standard gauge
transformations generated by the first-class constraints  and
their null-vectors (\ref{TRZ}):
\begin{equation}\label{g-tran}
\begin{array}{c}
    \delta_{\varepsilon} x^i=\{x^i,\Theta_I\}\varepsilon^I\,,\qquad \delta_{\varepsilon} \bar
    x_i = \{\bar x_i, \Theta_I\}\varepsilon^I\,,\\[5mm]
    \delta_{\varepsilon} \lambda^I=\dot{\varepsilon}{}^I -\lambda^K
    {U}_{KJ}^I\varepsilon^J +\Xi^I_A\varepsilon^A\,.
\end{array}
\end{equation}
Here $\varepsilon^I=(\varepsilon^a,\varepsilon^\alpha)$  and
$\varepsilon^A$ are infinitesimal gauge parameters, and the
structure functions ${U}_{KJ}^I(\phi)$ are defined by
(\ref{ThetaInv}).

Imposing the zero boundary conditions on the momenta
\begin{equation}\label{bcon}
    \bar x_i(t_1)=\bar x_i(t_2)=0\,,
\end{equation}
one can see \cite{LS2} that the classical dynamics of the model
(\ref{top}) are equivalent to those described by the original
(non-)Lagrangian equations $T_a=0$.

\vspace{3mm}\begin{example} Given the Lagrangian equations of motion
(\ref{dS}),  Lagrange anchor (\ref{V1}), and gauge symmetry
generators (\ref{RdS}), we have the following set of first-class
constraints on the phase space of fields and sources:
\begin{equation}\label{eTR}
    \widetilde{T}_i=\partial_iS +\bar x_i\,,\qquad
    \widetilde{R}_\alpha =R_\alpha^i\bar x_i\,.
\end{equation}
{}From the definition of gauge algebra it readily follows that
\begin{equation}\label{}
    \{\widetilde{T}_i,\widetilde{T}_j\}=0,\quad \{\widetilde{R}_\alpha,
    \widetilde{R}_\beta\}=U_{\alpha\beta}^\gamma \widetilde{R}_\gamma+
    U_{\alpha\beta}^{i}\widetilde{T}_i,\quad \{\widetilde{R}_\alpha,
    \widetilde{T}_i\}=U_{\alpha i}^j\widetilde{T}_j,
\end{equation}
where $U_{\alpha i}^j=\partial_i R^j_\alpha$ and
$U_{\alpha\beta}^i =\bar x_j W_{\alpha\beta}^{ij}(x)$.  Evidently,
the constraints (\ref{eTR}) are reducible,
\begin{equation}\label{}
    \widetilde{R}_\alpha =R^i_\alpha \widetilde{T}_i\,,
\end{equation}
and we can take $\{\widetilde{T}_i\}$ as a complete set of
independent first-class constraints. The corresponding Hamiltonian
action (\ref{top}) on the phase space of fields and sources reads
\begin{equation}\label{S_H}
    S_H[x,\bar x, \lambda ]=\int_{t_1}^{t_2} dt \left(\bar x_i\dot x^i - \lambda^i(\partial_iS(x)+\bar
    x_i)\right)\,.
\end{equation}
Excluding the momenta from this action by means of equations of
motion $\delta S/\delta \lambda^i=0$, we obtain
\begin{equation}\label{}
    S_H[x] =\int_{t_1}^{t_2} dt\, \dot
    x^i\partial_iS(x)=S(x(t_2))-S(x(t_1))\,.
\end{equation}
The latter action describes two copies of the original Lagrangian
theory corresponding to the ends of the ``time'' interval
$[t_1,t_2]$. As there is no coupling between the fields   $x(t_1)$
and $x(t_2)$, one can consistently restrict dynamics to either
subsystem with action $\pm S[x]$. This proves classical
equivalence of the topological theory with action (\ref{S_H}) to
(the two copies of) the Lagrangian theory with action $S[x]$.

\end{example}

\subsection{Physical observables}\label{PO}
The Poisson action of $\Omega$ on $\mathcal{N}$ makes the space
$C^\infty(\mathcal{N})$ into a cochain complex graded by ghost number: A
function $A$ is said to be BRST-closed if $\{\Omega, A\}=0$
and BRST-exact if $A=\{\Omega, B\}$ for some $B$.  Let
$H^n(\Omega)$ denote the corresponding cohomology groups. As
usual, the space of \textit{physical observables} is identified
with the group $H^0(\Omega)$, BRST cohomology at ghost number
zero.

It can be shown \cite{KLS} that the cohomology class of any BRST
cocycle $A$ with ghost number zero is completely determined by its restriction to
$M$, i.e., by the function $\bar{A}=A|_{M}$, and a function ${O}\in
C^{\infty}(M)$ is the restriction of some BRST cocycle iff
\begin{equation}\label{on-sell-inv}
    \langle R(\varepsilon),d{O}\rangle
    |_{\Sigma}=0\,,\qquad \forall\varepsilon \in \Gamma(\mathcal{F})\,.
\end{equation}
The trivial BRST cocycles are precisely those for which
${O}|_\Sigma =0$. Thus, to any on-shell gauge-invariant
function ${O}\in C^{\infty}(M)$ one can associate a BRST
cocycle and vice versa. Let $[A]\in H^0(\Omega)$ and $x_0\in
\Sigma$, then the map
\begin{equation}\label{hom}
    [A]\quad\mapsto \quad \langle A\rangle \equiv\bar A(x_0)\in
    \mathbb{R}
\end{equation}
establishes the isomorphism $H^0(\Omega)\simeq \mathbb{R}$. Since
$\Sigma\subset M$ is a connected submanifold and the distribution
$R$ acts on $\Sigma$ transitively (see Corollary  \ref{RS}), the map
(\ref{hom}) does not depend on the choice of $x_0\in \Sigma$. By
definition, $\langle A\rangle$ is the classical expectation value
of the physical observable $A$.

\section{Quantization}

In previous sections, we have described the procedure that assigns
a BRST complex to any dynamical system, be it Lagrangian or not.
The input data needed for constructing such a complex are the
classical equations of motion and the Lagrange structure. This
BRST complex has a clear physical interpretation as that resulting
from the BFV-BRST quantization of the topological sigma-model
(\ref{top}), whose target space is the cotangent bundle of the
space of all histories. By construction, the classical dynamics of
this effective topological theory are equivalent to the original
ones for any choice of the Lagrange structure. Quantizing now the
model (\ref{top}) by the usual BFV-BRST method, we induce some
quantization of the original (non-)Lagrangian theory; in so doing,
different Lagrange structures may result in different
quantizations of one and the same classical model.

Below, we start applying the standard prescriptions of the
BFV-BRST operator quantization to the constrained Hamiltonian
system (\ref{top}). What remains to specify is a convenient
representation. Here we prefer to work in the coordinate
(Schr\"odinger) representation, whereby a quantum state is
described by a wave-function on the ghost-extended space of all
histories. Then a physical wave-function is nothing but the
probability amplitude to find a system developing according to a
given history. For the Lagrangian systems, this amplitude is
simply given by the exponential of the action functional
multiplied by $i/\hbar$. In the non-Lagrangian case, however, it
may be a more general distribution, whose form strongly depends on
the choice of a Lagrange anchor. (see examples in Sec.6).

A consistent consideration of physical states in the coordinate
representation is known to require further enlargement of the
extended phase space by the so-called \textit{nonminimal
variables} \cite{HT}. These do not actually change the physical
content of the theory as one gauges them out by adding appropriate
terms to the original BRST charge. The nonminimal sector just
serves to bring the physical states to the ghost-number zero
subspace where one can endow them with a well-defined inner
product. We will not dwell on that in details, referring to the
textbook \cite{HT}. From now on, $\Omega$ will stand for the total
(i.e., nonminimal) BRST charge and the phase space $\mathcal{N}$
will include both minimal and nonminimal variables.

\subsection{Quantum BRST cohomology} Upon canonical quantization  each function on $\mathcal{N}$ turns to a linear
operator acting in a complex  Hilbert space $\mathcal{H}$:
\begin{equation}\label{}
    C^{\infty}(\mathcal{N}) \ni F\quad  \mapsto \quad \hat{F}\in
    \mathrm{End}(\mathcal{H})\,.
\end{equation}

A crucial step in the operator BFV-BRST quantization \cite{HT} is
assigning a nilpotent operator $\hat{\Omega}$ to the classical
BRST charge (\ref{Otheta}). The quantum symbol of the BRST
operator $\hat{\Omega}$ is supposed to have the form
\begin{equation}\label{symb}
    \Omega(\varphi,\bar\varphi, \hbar)=\sum_{k=0}^\infty
    \hbar^k\Omega^{(k)}(\varphi,\bar\varphi)\,,
\end{equation}
where the leading term $\Omega^{(0)}$ is given by  (\ref{Otheta})
and the higher orders in $\hbar$ are determined from the
requirements of hermiticity and nilpotency:
\begin{equation}\label{hn}
    \hat{\Omega}^\dagger=\hat{\Omega}\,,\qquad \hat{\Omega}^2=0\,.
\end{equation}
It may well happen that no $\hat{\Omega} $ exists satisfying these
two conditions, in which case one speaks about \textit{quantum
anomalies}. In what follows we assume our theory to be anomaly
free so that both equations (\ref{hn}) hold true.

In addition to the nilpotent BRST charge, the full BRST algebra
involves also  the anti-Hermitian ghost-number operator
$\hat{\mathcal{G}}$ such that $ [\hat{\mathcal{G}}, \hat{F}]=
\gh({F})\hat{F} $ for any homogeneous $\hat F$. In particular,
\begin{equation}\label{g-om}
    [\hat{\mathcal{G}},\hat\Omega]=\hat \Omega\,,\qquad
    \hat{\mathcal{G}}^\dagger=-\hat{\mathcal{G}}\,.
\end{equation}
Given the BRST algebra (\ref{hn}), (\ref{g-om}) one has two BRST
complexes.

The first one is given by the space of quantum state $\mathcal{H}$
with $\hat{\Omega}$ playing the role of coboundary operator. Under
certain assumptions \cite{HT} the space $\mathcal{H}$ splits as a
sum $\mathcal{H}=\oplus_{n\in \mathbb{Z}}\mathcal{H}^n$ of
eigenspaces of $\hat{\mathcal{G}}$ with definite real ghost
number. Then  $\hat{\Omega}:\mathcal{H}^n\rightarrow
\mathcal{H}^{n+1}$ is the cochain complex of quantum states graded
by ghost number. By $H_{\mathrm{st}}^n(\Omega)$ we denote the
$n$-th group of the BRST-state cohomology.

Associated with the BRST complex of quantum states is the complex
of quantum operators  $ \mathrm{End}(\mathcal{H})=\oplus_{n\in
\mathbb{Z}}\mathrm{End}^n(\mathcal{H})$. By definition,
$\hat{F}\in \mathrm{End}^n(\mathcal{H})$, iff
$\mathrm{ad}_{\hat{\mathcal{G}}}\hat F\equiv [\hat{\mathcal{G}},
\hat{F}]=n\hat{F}$. The corresponding coboundary operator
$\mathrm{ad}_{\hat{\Omega}}:
\mathrm{End}^n(\mathcal{H})\rightarrow
\mathrm{End}^{n+1}(\mathcal{H})$ acts by the rule
$\mathrm{ad}_{\hat{\Omega}}\hat{F}=[\hat\Omega, \hat{F}]$. The
$n$-th group of the BRST-operator cohomology is denoted by
$H_{\mathrm{op}}^n(\Omega)$.

The algebra of \textit{quantum physical observables} and the space
of \textit{quantum physical states} are then identified with the
corresponding BRST-cohomology at ghost number zero and the
physical dynamics are described in terms of the
$H^0_{\mathrm{op}}(\Omega)$-module $H^0_{\mathrm{st}}(\Omega)$.

Since the BRST charge is Hermitian, the inner product on
$\mathcal{H}$ induces that on the space of physical states
${H}^0_{\mathrm{op}}(\Omega)$. In many interesting cases, however,
the induced inner product appears to be ill defined  and needs
regularization. The most popular recipe to get a regular inner
product is to fold the BRST-closed operator $e^{\frac i\hbar
[\hat\Omega,\hat K]}\sim 1$ between a pair of BRST-closed states
$|\Psi_1\rangle,|\Psi_2\rangle\in\mathcal{H}$:
\begin{equation}\label{rinpr}
    \langle\Psi_1|\Psi_2\rangle_K=\langle\Psi_1|e^{\frac{i}{\hbar}[\hat\Omega,\hat
    K]}|\Psi_2\rangle\,,
\end{equation}
$K$ being an appropriate gauge-fixing fermion of ghost number
$-1$. Evidently, the last expression passes to the BRST cohomology
 and is independent  of a particular choice of $K$. (More precisely,
it depends only on the homotopy class of $K$ in the variety of all
gauge-fixing fermions providing finiteness of (\ref{rinpr}).) Now
the \textit{quantum average} of a physical observable
$[\hat{\mathcal{O}}]\in H_{\mathrm{op}}^0(\Omega)$ relative to a
physical $[|\Psi\rangle]\in H^0_{\mathrm{st}}(\Omega)$ is given by
\begin{equation}\label{QAVER}
    \langle\mathcal{O}\rangle=\frac{\langle\Psi|\hat{\mathcal{O}}|\Psi\rangle_K}{\langle\Psi|\Psi\rangle_K}\,.
\end{equation}

\subsection{Generalized Schwinger-Dyson equations}
As the Hamiltonian theory we deal with is topological, it might be
naively expected that $\dim_\mathbb{C}
H^0_{\mathrm{st}}(\Omega)=1$, so that the space of physical states
is spanned by a unique (up to equivalence) BRST-closed state
$|\Phi\rangle\in \mathcal{H}$. This would be quite natural because
the probability amplitude must be a unique distribution on the
space of all histories with prescribed boundary conditions.
Actually, it is not always the case in the BRST theory: The
physical dynamics may have several copies in the BRST-cohomology
(the $H^0_{\mathrm{op}}(\Omega)$-module
$H^0_{\mathrm{st}}(\Omega)$ is generally reducible), and choosing
one of them amounts to imposing extra conditions on the physical
states \cite[\S 14.2.6]{HT}. A guiding principle here is to
provide a positive-definiteness of the inner product (\ref{rinpr})
on a superselected physical space.

To be more specific, consider quantization of the gauge system
(\ref{brcov}), (\ref{Otheta}) in the case where $\mathcal{N}$ is a
superdomain endowed with canonical Poisson brackets, that is, in
formulas (\ref{brcov}), we just take $\nabla$ to be a flat
connection. Furthermore, we assume that the
$\varphi\bar\varphi$-symbol of the quantum BRST charge
(\ref{symb}) satisfies condition
\begin{equation}\label{}
    \Omega(\varphi,0,\hbar)=0\,.
\end{equation}
This property takes place for the leading (classical) term
$\Omega_0=\Omega(\varphi,\bar \varphi,0)$ and we require that it
holds true with account of all quantum corrections. Then, the
state $|\Phi\rangle$ that is annihilated by all the momenta,
\begin{equation}\label{Phi}
   \hat{ \bar\varphi}_I|\Phi\rangle=0\,,
\end{equation}
is annihilated by the BRST charge as well. After an appropriate
polarization of the nonminimal sector \cite{LS2}, the state
$|\Phi\rangle$ caries zero ghost number, and hence, defines a
physical state. In the coordinate representation, for instance, we
have $\Phi(\varphi)=c\in \mathbb{C}$. At first glance the
amplitude $\Phi$, being just a constant, has nothing to do with
the original dynamics, but that is illusion: The state
$|\Phi\rangle$ has an ill-defined norm in $\mathcal{H}$, so in
order to calculate the quantum average of a physical observable,
say $1\in H^0_{\mathrm{op}}(\Omega)$, one has use the regularized
inner product (\ref{rinpr}), but that brings an inevitable
dependence of the BRST charge. In other words, the information
about the original gauge system enters to the state $|\Phi\rangle$
implicitly, through passage to the BRST cohomology. To make this
dependence more explicit one should consider the BRST-dual of the
state $|\Phi\rangle$ (see \cite[\S 14.5.5]{HT} for general
definitions). The dual state looks like\footnote{Hereafter the
term ``ghosts'' refers to all fields with nonzero ghost number.}
\begin{equation}\label{}
    |\Phi'\rangle= |\psi\rangle\otimes|ghosts\rangle
\end{equation}
and is \textit{not} in general equivalent to the state
$|\Phi\rangle$. In the coordinate representation the first factor
$|\psi\rangle$, called the \textit{matter state}, is described by
a wave-function on $M$, while the second factor is given by a
wave-function of all other coordinates.  By definition, the matter
state is annihilated by the quantum constraints:
\begin{equation}\label{DQM}
\hat{\Theta}_I |\psi\rangle =0 \,,
\end{equation}
where the $x\bar x$-symbols of the constraint  operators
$\hat{\Theta}_I$ are given by
\begin{equation}\label{}
    \left. \Theta_I(x,\bar
x, \hbar)=\frac{\partial
\Omega(\varphi,\bar\varphi,\hbar)}{\partial
 \mathcal{C}^I}\right|_{ghosts=0}
\end{equation}
It is the state $|\psi\rangle$ that appears as physical state in
Dirac's quantization method. The consistency  of equations
(\ref{DQM}) implies the following commutation relations for the
quantum constraints:
\begin{equation}\label{}
    [\hat{\Theta}_I,\hat{\Theta}_J]=\hat{U}_{IJ}^K\hat{\Theta}_K
\end{equation}
with $\hat \Theta$'s to the right of $\hat U$'s. In view of the
last relation we can regard (\ref{DQM}) as a non-abelian
generalization of the Schwinger-Dyson equation to the case of
non-Lagrangian gauge theories. The next example justifies this
interpretation.

\vspace{3mm}
\begin{example}
Upon canonical quantization in the coordinate representation the
independent first-class constraints $\widetilde{T}_i$ in
(\ref{eTR}) turn to the pairwise commuting differential operators:
\begin{equation}\label{}
    \hat{\widetilde{T}}_i =\partial_i S(x)-i\hbar\frac{\partial}{\partial
    x^i}\,,\qquad
    [\hat{\widetilde{T}}_i,\hat{\widetilde{T}}_j]=0\,.
\end{equation}
Imposing these operators on the physical wave-function $\psi(x)$,
we arrive at the well-known Schwinger-Dyson equation in coordinate
representation
\begin{equation}
\left[\partial_i S(x)-i\hbar\frac{\partial}{\partial
    x^i}\right]\psi(x)=0           \,.
\end{equation}
A unique (up to an overall constant) solution to this equation is
given by the Feynman probability amplitude on $M$,
\begin{equation}\label{}
    \psi(x)=e^{-\frac i\hbar S(x)}\,.
\end{equation}
One can also quantize the constraints (\ref{eTR}) in the  momentum
representation, which is related to the coordinate one by the
(functional) Fourier transform.  The corresponding wave-function
\begin{equation}\label{}
    Z(\bar x)=\int Dx\, \psi(x)e^{-\frac{i}{\hbar}x^i\bar x_i}=\int Dx\, e^{-\frac{i}{\hbar}(S(x)+x^i\bar x_i)}
\end{equation}
is nothing but the generating functional of Green's functions with
$\bar x$'s playing the role of classical sources.
\end{example}
\vspace{3mm}

In principle, one can use any copy of a single physical state in
the BRST-state cohomology to compute  the quantum average of a
physical observable $[\mathcal{O}]\in H^0_{\mathrm{op}}(\Omega)$
by formula (\ref{QAVER}). It is also possible and is particularly
convenient to use the asymmetric definition for the quantum
averages:
\begin{equation}\label{expect}
\langle{\mathcal{O}}\rangle=\frac{\langle
\Phi'|\hat{\mathcal{O}}|\Phi\rangle_K}{\langle \Phi'
|\Phi\rangle_K}
\end{equation}
In  \cite{LS2}, it was shown that all such definitions give one
and the same value $\langle{\mathcal{O}}\rangle$. Similarly to the
BRST-state cohomology, the BRST-operator cohomology is essentially
one-dimensional that allows one to establish a one-to-one
correspondence between the physical states and physical
observables. Namely, given a physical observable
$\hat{\mathcal{O}}$, we define the physical state
\begin{equation}\label{}
|O\rangle =\hat{\mathcal{O}}|\Phi\rangle \,.
\end{equation}
The latter is necessarily of the form $|O\rangle =\langle
{\mathcal{O}}\rangle|\Phi\rangle + \hat{\Omega}|\Lambda\rangle$.
Using  the coordinate representation, we can rewrite
(\ref{expect})  as
\begin{equation}\label{}
\langle{\mathcal{O}}\rangle=\frac{\langle
\Phi'|O\rangle_K}{\langle \Phi'|1\rangle_K} =(\mathrm{const})\int
D\varphi O(\varphi)\Phi'_K(\varphi)\,.
\end{equation}
The last expression enables us to treat the gauge-fixed
probability amplitude $\Phi'_K(\varphi)=\langle
\Phi'|\varphi\rangle_K$ as a linear functional on the space of
physical observables represented by the physical states
$O(\varphi)=\langle\varphi|O\rangle$.

\subsection{Path-integral representation} Regarding the regulator
$e^{\frac i\hbar [\hat{\Omega},\hat{K}]}$ in (\ref{rinpr}) as the
evolution operator corresponding to the BRST-trivial Hamiltonian
$\hat H= [\hat\Omega,\hat K]$, we can immediately write down the
path-integral representation for the quantum average
(\ref{QAVER}):
\begin{equation}\label{Qav}
\begin{array}{c}
\displaystyle\langle \mathcal{O} \rangle=\frac{\langle \Phi|
\hat{\mathcal{O}}e^{\frac i\hbar [\hat{\Omega},\hat{K}]}
|\Phi\rangle}{\langle \Phi|e^{\frac i\hbar
[\hat{\Omega},\hat{K}]}|\Phi\rangle}=\\[7mm]
\displaystyle=(\mathrm{const})\int \mathcal{D}\varphi
    \mathcal{D}\bar\varphi\, O(\varphi(1)) \exp{\frac i\hbar\int_0^1 dt (\bar\varphi_I\dot{\varphi}^I -
    \{\Omega,K\})}\,.
    \end{array}
\end{equation}
Here the normalization constant is chosen in such a way that
$\langle 1\rangle=1$ and integration extends over all fields
obeying
\begin{equation}\label{BC}
    \bar\varphi_I(0)=\bar\varphi_I(1)=0\,.
\end{equation}
These boundary conditions follow directly from definition
(\ref{Phi}) of the physical state $|\Phi\rangle $. Because of
(\ref{BC}) only the $\bar\varphi$-independent part
\begin{equation}\label{O}
O(\varphi)=\mathcal{O}(\varphi,\bar\varphi)|_{\bar\varphi
=0}={O}(x) + (ghost \;terms)
\end{equation}
of the physical observable $\mathcal{O}$ contributes to the path
integral (\ref{Qav}). It is not hard to see that the function
(\ref{O}) obeys to (and can be determined from) the following
equation:
\begin{equation}\label{}
Q O =\{\Omega_1, O\}=0\,,
\end{equation}
$Q$ being the classical BRST differential (\ref{Q}).

Note that (\ref{Qav}) is nothing but the usual Feynman's path
integral for the topological sigma-model with action
\begin{equation}\label{S}
S[\varphi,\bar\varphi]=\int_0^1 dt (\bar\varphi_I\dot{\varphi}^I -
    \{\Omega,K\})\,.
\end{equation}
This can be viewed as resulting from the BFV quantization of the
constrained Hamiltonian theory (\ref{top}). If $\dim X=d$, where
$X$ is the initial space-time manifold, then (\ref{S}) defines a
topological field theory on the $(d+1)$-dimensional manifold
$\widetilde{X}=X\times I$ with boundary. Suppose $X$ is an
orientable manifold, then so is $\widetilde{X}=X\times I$ and each
of the two orientations of $\widetilde{X}$ induces opposite
orientations on the connected components of the boundary $\partial
\widetilde{X}=X_0\cup X_1$; here $X_0 \simeq X\simeq X_1$ and the
subscripts $0$ and $1$ refer to the different orientations of $X$.

As the model (\ref{S}) is purely  topological, there are no
physical dynamics in the bulk of $\widetilde{X}$. Put differently,
all the physical degrees of freedom, if any, are supported at the
boundary $\partial \widetilde{X}=X_0\cup X_1$, where they evolve
according to the classical equations of motion $T_a=0$; in so
doing, the dynamics on $X_0$ and $X_1$ are completely independent
of each other. Thus, the action (\ref{S}) describes two copies of
the same filed-theoretical model, which defer only by orientation
of the space-time manifold $X$ (two parallel universes).

This classical consideration can be further promoted to the
quantum-mechanical level. Consider the \textit{projected kernel}
associated with the matter state (\ref{DQM}). It can be defined by
the path integral \cite{HT}:
\begin{equation}\label{}
    \psi(x_1)\bar\psi(x_0)=\int D\varphi D\bar\varphi\, e^{\frac i\hbar
    S[\varphi,\bar\varphi]}\,,
\end{equation}
where the sum runs over trajectories
$(\varphi^I(t),\bar\varphi_J(t))$ subject to appropriate boundary
conditions at $t=0,1$. In particular, $x_1=x(1)$, $x_0=x(0)$; the
boundary conditions for the other variables can be found in
\cite{LS2}. According to our definitions, the state $\psi$
describes a gauge invariant probability amplitude for a field
theory on $X_0$. Then $\bar\psi$ must play the same role for
$X_1$. \footnote{In the Lagrangian field theory, for example, the
probability amplitude has the form $\psi=e^{\frac i\hbar S}$,
where the action functional is  given by the integral $S=\int_X L$
of some top form $L$ (a Lagrangian density). Changing an
orientation of $X$ yields $S\rightarrow -S$, hence
$\psi\rightarrow \bar\psi$.} Multiplying $\psi$ by $\bar\psi$, we
get the right probability amplitude for the field theory on
$X_0\cup X_1$, as there is no interaction between the fields on
$X_0$ and $X_1$ (correlations through the bulk of $\widetilde{X}$
are completely suppressed  by gauge invariance).

\vspace{3mm}
\begin{example} Let us compute the quantum average (\ref{Qav}) for the
topological model (\ref{S_H}), where the action $S$ is not gauge
invariant. A good gauge-fixing condition in the bulk is the
derivative gauge
\begin{equation}\label{}
    \ddot x^i =0\,.
\end{equation}
As with any abelian gauge theory, the ghost fields are decoupled
from the matter ones and can thus be integrated out explicitly.
The result is given by
\begin{equation}\label{INT}
\begin{array}{c}
\displaystyle \langle \mathcal{O}\rangle= c\int DxD\bar x D\lambda
\delta[\ddot x] O(x(1))e^{\frac i\hbar S_H[x,\bar x,\lambda]}\\[5mm]
\displaystyle=c\int Dx \delta[\ddot x]O(x(1))\exp \frac i\hbar
\left(S[x(1)]-S[x(0)]\right)\\[5mm]
\displaystyle = c'\int Dx(1) O(x(1)) e^{\frac i\hbar
S[x(1)]}=\langle O|\psi \rangle \langle \psi | 1\rangle\,,
\end{array}
\end{equation}
where
\begin{equation}\label{}
\psi (x)=e^{\frac i\hbar S[x]}\,,\qquad    c{}'= c\langle \psi|
1\rangle=c\int Dx(0)e^{-\frac i\hbar S[x(0)]}\,.
\end{equation}
So, up to a normalization constant, the integral (\ref{INT}) gives
the usual quantum average of an observable $O$ in the Lagrangian
theory with action $S$. Notice that one can arrive at the same
result by imposing the derivative gauge on the Lagrange multiplier
$\dot \lambda^i=0$.

\end{example}

\section{Augmentation}

In previous sections, we have formulated the quantization
procedure for (non-)Lagrangian gauge theories, which starts with
the classical equations of motion and Lagrange structure as input
data and results in the generalized Schwinger-Dyson equation for
the probability amplitude on the space of all histories $M$. We
have also seen that the amplitude  admits a simple path-integral
representation in terms of a Lagrangian topological field theory
in the space-time with one more dimension. In this section, we
derive an alternative path-integral representation for the
probability amplitude of a (non-)Lagrangian theory in terms of
some Lagrangian model on the same space-time manifold, but
augmented with extra fields. The configuration space of the
augmented field theory is taken to be the total space of the
vector  bundle ${\mathcal E}^\ast\rightarrow M$, the dual to the
dynamics bundle $\mathcal{E}$; in so doing, the original
configuration space $M$ is embedded in $\mathcal E^\ast$ as the
zero section. The augmentation procedure extends the original
(non-)Lagrangian dynamics from $M$ to $\mathcal{E}^\ast$ in such a
way that the entire system becomes Lagrangian. We show that, at
classical level, the augmented theory is equivalent to the
original one provided that special boundary conditions are fixed
for the augmentation fields. At quantum level, integrating the
Feynman probability amplitude on $\mathcal{E}^\ast$  over the
augmentation fields yields the probability  amplitude on $M$.

\subsection{An augmented BRST complex. }
Augmentation of the original dynamics on $M$  implies a consistent
extension to $\mathcal{E}^\ast$ of the original equations of
motion, gauge symmetries, Noether identities, and the Lagrange
structure. As a practical matter, it is convenient to make these
extensions not at the level of the space of all histories, but
augmenting the ambient symplectic manifold $\mathcal{N}$, which
already involves all necessary ghost fields of the original
theory. The overall result of these extensions turns out to be
just a ``duplication'' of the ambient manifold. More precisely,
the manifold $\mathcal{N}$, considered as the total space of the
vector bundle (\ref{N}), is replaced with
$\mathcal{N}_{\mathrm{aug}}=\mathcal{N} \oplus \Pi(\mathcal{N}
\oplus TM)$. The table below contains the data on various gradings
assigned to the fiber coordinates of $\Pi(\mathcal{N}\oplus TM)$:

$$
\begin{tabular}{|l|c|c|c|c|c|c|c|c|}
  \hline
fibers           &$\mathcal{E}^\ast$ & $\mathcal{E}$ &
$\mathcal{G}^\ast $&$\mathcal{G}$ &$T^\ast M$&$
T M$&$\mathcal{F}^\ast$& $\mathcal{F}$\\
\hline fiber coordinates          &\,$y^a$ & $\bar y_b$&\, $c^A
$\,& \,$\bar c_{B}$\, &\, $\eta_i$ & \,$\bar
\eta^j$&\,$\xi_\alpha$\,&\,$\stackrel{}{\bar
           \xi^\beta}$\\
  \hline
  $\epsilon$=Grassman parity            & 0 & 0 & 1 & 1 & 1 & 1 & 0 & 0\\
  \hline
  $\mathrm{gh}$ = ghost number                   & 0 & 0 & 1 & -1 & -1 & 1 & -2&2\\
  \hline
  $\mathrm{Deg}$ = momentum degree      & 0 & 1 & 0 & 1 & 0 & 1&0&1 \\
  \hline
  $\deg$ = resolution degree               & 1 & 0 & 1 & 0 & 1 & 0& 1& 0\\
  \hline
\end{tabular}
$$
\begin{center}
{Table 2}
\end{center}

In order to compare the ghost numbers of the new and old fields it
is convenient to assemble the augmentation fields into ``position
coordinates'' and ``momenta'':
\begin{equation}
\varphi^{\mathrm{aug}}_I=(\eta_i, \xi_\alpha, y^a, c^A)\,,\qquad
\bar\varphi^I_{\mathrm{aug}}=(\bar\eta^i, \bar\xi^\alpha, \bar
y_a, \bar c_A)\,.
\end{equation}
Then we have
\begin{equation}
\begin{array}{ll}
\mathrm{gh}(\varphi_I^{\mathrm{aug}})=-\mathrm{gh}(\bar\varphi^I_{\mathrm{aug}})\,,\qquad
&\epsilon(\varphi_I^\mathrm{aug})=\epsilon(\bar\varphi^I_{\mathrm{aug}})\,,\\[5mm]
\mathrm{Deg}(\bar\varphi^I_\mathrm {aug})=1\,,\qquad&
\mathrm{Deg}(\varphi_I^\mathrm {aug})=0\,,
\end{array}
\end{equation}
and
\begin{equation}\label{}
    \mathrm{gh}(\varphi_I^\mathrm{aug})=\mathrm{gh}(\bar\varphi_I)-1\,,\qquad
    \mathrm{gh}(\bar\varphi^I_\mathrm{aug})=\mathrm{gh}(\varphi^I)+1\,.
\end{equation}
So, the ``duplication'' of the ambient manifold $\mathcal{N}$ is
accompanied with reversion of parities and shift of ghost numbers.

As a next step, we extend the exact symplectic structure
(\ref{Theta}) on $\mathcal N$ to that on ${\mathcal
N}_{\mathrm{aug}}$ by setting
\begin{equation}\label{om-aug}
\omega_{\mathrm {aug}} =\omega + d(\bar \varphi^I_{\mathrm
{aug}}\nabla \varphi_I^{\mathrm {aug}})\,,
\end{equation}
$\nabla$ being some connection on ${\mathcal N}\oplus TM$.

Finally,  the original BRST charge $\Omega$ on $\mathcal{N}$ is
extended to $ \mathcal{N}_{\mathrm{aug}}$ as
\begin{equation}\label{augBRST}
    \Omega_{\mathrm {aug}} =\Omega +\sum_{n=0}^\infty \Omega_n\,,\qquad
    \deg(\Omega_n)=n\,.
\end{equation}
Here
\begin{equation}\label{}
    \Omega_0= \bar\varphi_I\bar\varphi_{\mathrm {aug}}^I
\end{equation}
and the higher orders in the resolution degree are determined from
the master equation
\begin{equation}\label{mea}
    \{ \Omega_{\mathrm {aug}}, \Omega_{\mathrm {aug}}\}=0\,.
\end{equation}

Let us show that the last equation has a solution indeed. To this
end, we introduce the following  pair of nilpotent operators:
\begin{equation}
\begin{array}{lll}
\displaystyle\delta =\bar\varphi_I\frac{\partial}{\partial \varphi_I^{\mathrm{aug}}} \,,\qquad &
    \delta^2=0\,,\qquad & \deg(\delta)=-1\,,\\[7mm]
\displaystyle \delta^\ast = \varphi^{\mathrm {aug}}_I\frac{\partial}{\partial \bar\varphi_I}\,,\qquad &
(\delta^\ast)^2=0\,,\qquad &\deg(\delta^\ast)=1\,.
\end{array}
\end{equation}
It is straightforward  to check that
\begin{equation}\label{Hodge}
    \delta\delta^\ast+\delta^\ast\delta=N\,, \qquad N\equiv N_1+N_2\,,
\end{equation}
where the operator
\begin{equation}\label{N1}
    N_1= {\bar\varphi_I}\frac{\partial}{\partial \bar\varphi_I}
\end{equation}
counts the momentum degree of the ``old" variables (see Table 1),
while
\begin{equation}\label{}
    N_2=\varphi^{\mathrm {aug}}_I\frac{\partial}{\partial \varphi^{\mathrm{aug}}_I}
\end{equation}
counts the resolution degree of augmentation fields. If we regard
$\delta$ as the differential of the cochain complex
$C^{\infty}({\mathcal{N}_{\mathrm{aug}}})$, then $\delta^\ast$
becomes a homotopy for $N$ with respect to $\delta$. As a result,
all the nontrivial $\delta$-cocycles are nested in the subspace
$\ker \, N\subset C^\infty({\mathcal{N}_{\mathrm{aug}}})$.

Now applying the standard technique of homological perturbation
theory \cite{HT}, we can prove the following statement.

\begin{proposition}
There is a unique BRST charge (\ref{augBRST}) satisfying the
master equation (\ref{mea}) and the condition
\begin{equation}\label{adcon}
    \delta^\ast(\Omega_{\mathrm{aug}}-\Omega-\Omega_0)=0\,.
\end{equation}
\end{proposition}

\prf\, Expanding the master equation (\ref{mea}) with respect to
the resolution degree, we arrive at the following sequence of
equations:
\begin{equation}\label{B}
    \delta \Omega_{n+1}=B_n(\Omega_0,...,\Omega_n)\,,\qquad
   n\in \mathbb{N}\,,
\end{equation}
where
\begin{equation}
\displaystyle B_n =  P_n\Big( \{\Omega,\Omega_n\}+
\sum_{s=0}^n\{\Omega_{n-s},\Omega_s\}\Big)\,,
\end{equation}
and $P_n$ is the projector on the subspace of functions of
resolution degree $n$. We can solve these equations in series
starting with $\delta\Omega_1=B_0$. Since $\deg B_n =n$ and the
function $B_0=\{\Omega,\Omega_0\}$ contains no terms of zero
momentum degree w.r.t. the old variables, the operator $N$ is
invertible on the subspace $W\subset
C^{\infty}(\mathcal{N}_{\mathrm{aug}})$ spanned by all $B$'s. The
condition $\delta B_n=0$ is then necessary and sufficient for the
$n$-th equation (\ref{B}) to be solvable. The closedness of $B_n$
is established by induction on $n$, just putting successive
restrictions on the resolution degree of the Jacobi identity
$\{\Omega,\{\Omega,\Omega\}\}\equiv 0$.

Finally, applying the operator $\delta^\ast$ to both sides of Eq.
(\ref{B}) and using Rels. (\ref{Hodge}), (\ref{adcon}), we get the
following recurrent relations for the homogeneous components of
$\Omega$:
\begin{equation}\label{rec-rel}
\Omega_{n+1}= \delta^\ast (N|_W)^{-1}B_n(\Omega_0,...,\Omega_n)\,.
\end{equation}
Here we have used the fact that the operator $\delta^\ast$
commutes with $N$ and, as a consequence, with $(N_W)^{-1}$. By
construction, $\delta^\ast\Omega_n=0$, $\forall n> 0$, so that the
augmented BRST charge $\Omega_{\mathrm{aug}}$ meets equation
(\ref{adcon}). $\square$

In sequel we will need the following property of the augmented
BRST charge.

\begin{proposition}\label{52}
If the augmented BRST charge (\ref{augBRST}) satisfies
(\ref{adcon}), then
\begin{equation}\label{cond}
(\Omega_{\mathrm {aug}} - \Omega-\Omega_0)|_{\bar
\varphi^I_{\mathrm {aug}} =0}=0\,.
\end{equation}
\end{proposition}

\prf \,  by induction on resolution degree.

\subsection{Interpretation} Given the augmented BRST charge (\ref{augBRST}),
one may ask what is a classical theory this BRST charge
corresponds to (or can be derived from).  According to the general
definitions of Sec. 3, the equations of motion,  gauge symmetry
and Noether identity generators, as well as the Lagrange anchor,
enter the BRST charge  as coefficients at lower powers  of  fiber
coordinates with certain ghost numbers and momentum degrees (see
relation (ii) at the beginning of Sec.3.2). So, to identify all
the key ingredients of the underlying gauge dynamics we are going
just to evaluate the appropriate terms in the augmented BRST
charge. Thus, the equations of motion define the terms that are
linear in fiber coordinates:
\begin{equation}\label{T-aug}
\displaystyle    \mathbf{T}_a=
\left.\frac{\partial\Omega_{\mathrm{aug}}}{\partial \bar
\eta^a}\right|_{\mathcal{E}^\ast}=T_a(x)=0\,,\qquad
\mathbf{T}_i=\left.\frac{\partial \Omega_{\mathrm {aug}}}{\partial
\bar\eta^i}\right|_{\mathcal{E}^\ast}=\nabla_i T_a y^a +
o(y^2)=0\,.
\end{equation}
As is seen, the first group of equations coincides with the
original equations of motion  on $M$. The absence of
$y$-contributions to these equations is guaranteed by Proposition
\ref{52}. So, the original dynamics on $M$ are completely
decoupled from the augmented system (\ref{T-aug}). The second
group of equations, being at least linear in $y$'s, admit a
trivial solution $y^a=0$, which can be singled out by imposing
zero boundary conditions on $y$'s.

In general, the augmented equations of motion (\ref{T-aug}) are
both gauge invariant and linearly dependent. It follows from
definiens of Sec.3.2 that the  gauge algebra generators are given
by
\begin{equation}\label{R-aug}
\begin{array}{l}
    \displaystyle \mathbf{R}_\alpha =\left. \frac{\partial^2 \Omega_{\mathrm{aug}}}{\partial c^\alpha \partial \bar x_i}\right |_{\mathcal{E}^\ast}\frac{\partial}{\partial x^i}+ \left. \frac{\partial^2 \Omega_{\mathrm{aug}}}{\partial c^\alpha \partial \bar y_a}\right |_{\mathcal{E}^\ast}\frac{\partial}{\partial y^a}=R_\alpha^i(x)\frac{\partial}{\partial x^i}
    +o(y)\,,\\[7mm]
   \displaystyle \mathbf{R}_A= \left. \frac{\partial^2 \Omega_{\mathrm {aug}}}{\partial c^A \partial \bar y_a}\right |_{\mathcal{E}^\ast}\frac{\partial}{\partial y^a}  + \left. \frac{\partial^2 \Omega_{\mathrm {aug}}}{\partial c^A \partial \bar x_i}\right |_{\mathcal{E}^\ast}\frac{\partial}{\partial x^i} = Z^a_A(x)\frac{\partial}{\partial
    y^a}+ o(y)\,,
    \end{array}
\end{equation}
so that
\begin{equation}\label{Transf-aug}
    \begin{array}{ll}
    \mathbf{R}_\alpha
    \mathbf{T}_a=\mathbf{U}_{\alpha a}^b\mathbf{T}_b+\mathbf{U}_{\alpha a}^i\mathbf{T}_i\,,\qquad &
    \mathbf{R}_A
    \mathbf{T}_a=\mathbf{U}_{A a}^b\mathbf{T}_b+\mathbf{U}_{A a}^i\mathbf{T}_i\,,\\[7mm]
    \mathbf{R}_\alpha
    \mathbf{T}_i=\mathbf{U}_{\alpha i}^b\mathbf{T}_b+\mathbf{U}_{\alpha i}^j\mathbf{T}_j\,,\qquad &
    \mathbf{R}_A
    \mathbf{T}_i=\mathbf{U}_{A i}^b\mathbf{T}_b+\mathbf{U}_{A i}^j\mathbf{T}_j\,,
    \end{array}
\end{equation}
for some structure functions $\mathbf{U}$.  The Noether identities
have the following form in the augmented theory:
\begin{equation}\label{Ident-aug}
    \mathbf{Z}^a_A
    \mathbf{T}_a+\mathbf{Z}^i_A\mathbf{T}_i = 0\,,\qquad
    \mathbf{Z}^a_\alpha
    \mathbf{T}_a+\mathbf{Z}^i_\alpha \mathbf{T}_i = 0\,,
\end{equation}
where
\begin{equation}\label{Z-aug}
\begin{array}{ll}
    \displaystyle \left. \mathbf{Z}^a_A=\frac{\partial^2 \Omega_{\mathrm{aug}}}{\partial \bar \xi^A\partial \eta_a}\right |_{\mathcal{E}^\ast}=Z^a_A(x)+o(y)\,,\qquad&\displaystyle
    \mathbf{Z}_A^i=\left.\frac{\partial^2\Omega_{\mathrm{aug}}}{\partial \bar\xi^A\partial \eta_i }\right |_{\mathcal{E}^\ast}=o(y)\,,\\[7mm]

    \displaystyle \displaystyle
    \left.\mathbf{Z}^i_\alpha=\frac{\partial^2 \Omega_{\mathrm{aug}}}{\partial \bar \xi^\alpha \partial \eta_i }\right|_{\mathcal{E}^\ast}
    =R_\alpha^i(x)+o(y)\,, &\displaystyle\left.
     \mathbf{Z}^a_\alpha=\frac{\partial^2 \Omega_{\mathrm{aug}}}{\partial \bar \xi^\alpha \partial\eta_a}\right|_{\mathcal{E}^\ast}=o(y)\,.
    \end{array}
\end{equation}

As is seen from Rels. (\ref{R-aug}), there are two types of gauge
symmetry transformations in the augmented theory. The first ones,
generated by $\mathbf{R}_\alpha$, are just extensions to
$\mathcal{E}^\ast$ of the original gauge symmetries. The second
type transformations, generated by $\mathbf{R}_A$, start from the
vertical vector fields on $\mathcal{E}^\ast$ associated with the
Noether identity generators $Z_A^a$.

Looking at the generators of Noether's identities (\ref{Z-aug}),
one can observe a mirror inversion in the structure of the gauge
symmetry generators: The generators  $R_\alpha$ on $M$ give rise
to the Noether identity generators $\mathbf{Z}_\alpha$, while the
generator $\mathbf{Z}_A$ is just a continuation of corresponding
Noether identity generator from the original theory.

We thus conclude that the numbers of Noether's identities and
gauge symmetries coincide in the augmented theory. Furthermore,
the expansion in the augmentation fields $y^i$ starts with the
same terms for both sets of generators. And one can further deduce
that the generators of identities (\ref{Z-aug}) coincide with the
generators of gauge symmetries (\ref{R-aug}). Such a pairing
between Noether identities and gauge symmetries is characteristic
for the Lagrangian dynamics.

To further elucidate  the meaning  of the augmented BRST charge in
terms of the phase space of fields $x^i$, $y^a$ and their sources
$\bar x_i$, $\bar y_a$, we introduce the following collective
notation:
\begin{equation}
\phi^{\bar a}=(x^i, y^a)\,,\qquad \bar\phi_{\bar a} =(\bar x_i,
\bar y_a) \,, \qquad \bar\eta^{\bar a}=(\bar\eta^i, \bar\eta^a)\,.
\end{equation}
Then the deformed phase-space constraints associated with the
augmented equations of motion (\ref{T-aug})  are given by
\begin{equation}\label{TT-aug}
\begin{array}{c}
\displaystyle    \widetilde {\mathbf T}_{\bar a} =
    \left.\frac{\partial\Omega_{\mathrm{aug}}}{\partial \bar\eta^{\bar a}}\right|_{{\mathcal E}^\ast\oplus T^\ast M\oplus \,{\mathcal E}} \, = \\[7mm]
 \displaystyle   =\mathbf{T}_{\bar a}(\phi)+ V_{\bar a}^{\bar b}(\phi)\bar\phi_{\bar b}+\sum_{k=2}^\infty V_{\bar a}^{{\bar b}_1\cdots
    {\bar b}_k}(\phi)\bar \phi_{{\bar b}_1}\cdots\bar\phi_{{\bar b}_k}\approx 0\,.
    \end{array}
\end{equation}

According to our definitions, the coefficients $V_{\bar a}^{\bar
b}(\phi)$ in (\ref{TT-aug}) are to be identified with the
components of the Lagrange anchor. Using the recurrent relations
(\ref{rec-rel}), we find
\begin{equation}\label{V}
V=(V^{\bar b}_{\bar a})=\left(
\begin{array}{cc}
  V_a^i(x) & \delta_a^b \\[3mm]
  \delta^i_j & 0 \\
\end{array}%
\right)+ o(y)\,.
\end{equation}
As is seen the augmented Lagrange anchor is always nondegenerate
and its inverse has the form
\begin{equation}\label{inv1}
    \Lambda =V^{-1}=\left(%
\begin{array}{cc}
  0 & \delta^i_j \\[3mm]
  \delta_a^b & -V^i_a(x) \\
\end{array}%
\right)+o(y)\,.
\end{equation}

To make contact with the definitions of Sec.2, we identify the
total space of the tangent bundle $T{\mathcal E}^\ast$ with the
total space of ${\mathcal E}^\ast\oplus TM\oplus {\mathcal
E}^\ast$ and the total space of $T^\ast{\mathcal E}^\ast$  with
that of ${\mathcal E}\oplus T^\ast M\oplus {\mathcal E}^\ast$ by
making use the linear connection $\nabla$ on
$\mathcal{E}^\ast\rightarrow M$. Upon these identifications, the
bundle map $T\mathcal{E}^\ast\rightarrow \mathcal{E}^\ast$ goes
into the bundle map ${\mathcal E}^\ast\oplus TM\oplus {\mathcal
E}^\ast\rightarrow \mathcal{E}^\ast$ (projection on the third
factor) and the same is true for the cotangent bundle $T^\ast
\mathcal{E}^\ast$. Now we can summarize the discussion above as
follows.

\begin{proposition}
The augmented BRST complex describes a complete Lagrange structure
of type (1,1) associated to the on-shell exact sequence
\begin{equation}
0\rightarrow {\mathcal V}\rightarrow T{\mathcal E}^\ast\rightarrow
T^\ast{\mathcal E}^\ast\rightarrow {\mathcal V}\rightarrow 0\,,
\end{equation}
where the gauge algebra (= Noether identity) bundle $ {\mathcal
V}$ is the vector bundle with the base ${\mathcal E}^\ast$, total
space ${\mathcal F}\oplus {\mathcal E}^\ast\oplus {\mathcal G}$,
and the bundle map $ p: {\mathcal F}\oplus {\mathcal E}^\ast\oplus
{\mathcal G}\rightarrow {\mathcal E}^\ast$ (projection on the
second factor).
\end{proposition}

The completeness of the augmented Lagrange structure has two
immediate consequences. First of all, the constraints
(\ref{TT-aug}) define a Lagrangian submanifold in the augmented
phase space ${\mathcal E}^\ast \oplus T^\ast M\oplus{\mathcal E}$,
so that the rest of the constraints, namely, the constraints
\begin{equation}\label{RR-aug}
\begin{array}{l}
    \displaystyle \widetilde{\mathbf{R}}_\alpha =\left. \frac{\partial \Omega_{\mathrm{aug}}}{\partial c^\alpha}\right
    |_{\mathcal{E}^\ast\oplus \,T^\ast M\oplus\, \mathcal{E}}=\widetilde{R}_\alpha (x,\bar x)
    +o(y)\,,\\[7mm]
   \displaystyle \widetilde{\mathbf{R}}_A= \left. \frac{\partial \Omega_{\mathrm {aug}}}{\partial c^A}\right |_{\mathcal{E}^\ast\oplus\,
   T^\ast M\oplus\,\mathcal{E}}= \widetilde{Z}^a_A(x, \bar x)
    \bar y_a+ o(y)
    \end{array}
\end{equation}
associated with the gauge symmetry generators (\ref{R-aug}),
(\ref{TRZ}), are given by  linear combinations of (\ref{TT-aug}).
The second consequence is that, according to Theorem \ref{LA}, the
augmented equations of motion (\ref{T-aug}) are equivalent to
Lagrangian ones.

To get an explicit expression for corresponding action functional,
one has just to resolve the first-class constraints (\ref{TT-aug})
with respect to momenta $\bar\phi_{\bar a}$. This can always be
done at least perturbatively. As a starting point, we rewrite the
constraint equations (\ref{TT-aug}) in the following equivalent
form:
\begin{equation}\label{inv2}
\bar\phi_{\bar a}=-\Lambda_{\bar a}^{\bar b} {\mathbf T}_{\bar b}
- \Lambda_{\bar a}^{\bar b} \sum_{k=2}^\infty V_{\bar b}^{{\bar
a}_1\cdots {\bar a}_k}\bar \phi_{{\bar a}_1}\cdots \bar\phi_{{\bar
a}_k}\,  ,
\end{equation}
where $\Lambda$ is defined by relation (\ref{inv1}). Then, taking
$\bar\phi=-\Lambda{\mathbf T}$ as zero order approximation and
iterating these equations ones and again, we finally arrive at the
equivalent set of first-class constraints
\begin{equation}\label{T'-aug}
\widetilde{\mathbf T}'_{\bar a}=\bar\phi_{\bar a}-{\mathbf
T}'_{\bar a}(\phi)\approx 0\,,
\end{equation}
where
\begin{equation}
{\mathbf T}'_{\bar a} = \sum_{k=1}^\infty F^{\bar b_1\cdots \bar
b_k}(\phi) {\mathbf T}_{{\bar b}_1}(\phi)\cdots {\mathbf T}_{{\bar
b}_k}(\phi) = \widetilde\Lambda(\phi)^{\bar b}_{\bar a} {\mathbf
T}_{\bar b}(\phi)\,.
\end{equation}
The constraints (\ref{T'-aug}), being resolved w.r.t. momenta
$\bar\phi_{\bar a}$, are to be necessarily commuting,
\begin{equation}
\{\widetilde{\mathbf T}'_{\bar a}, \widetilde{\mathbf T}'_{\bar
b}\}=0\,,
\end{equation}
that amounts to existence of an action functional $S(\phi)$ such
that
\begin{equation}\label{S'}
 {\mathbf T}'_{\bar a}(\phi)=\partial_{\bar a} S(\phi)\,.
\end{equation}
Thus, the  augmented equations of motion are equivalent to the
Lagrangian equations (\ref{S'}) with $\widetilde \Lambda $ playing
the role of integrating multiplier. Finally, using the standard
homotopy operator for the exterior differential, we can
reconstruct the action as
\begin{equation}\label{Shom}
S(\phi) = \phi^{\bar a} \int_0^1 {\mathbf T}'_{\bar a}(s\phi) ds +
(\mathrm{const})\,.
\end{equation}
Up to the second order in $y$'s and an inessential additive
constant the action reads
\begin{equation}\label{Sxy}
S(x,y)= T_a(x)y^a+G_{ab}(x)y^ay^b + o(y^3)\,.
\end{equation}
Here the symmetric matrix
\begin{equation} \label{VV}
G_{ab} = V_a^i\nabla_i T_b+V_b^i\nabla_i T_a
\end{equation}
can be thought of as a generalization of  Van Vleck's matrix. It
is the matrix that defines the form of the first quantum
correction to the classical average of physical observables
\cite{KLS}.

More explicitly,  the equations of motion (\ref{S'}) following
from variation of (\ref{Shom}) read
\begin{equation}\label{varS}
    \mathbf{T}'_a=\frac{\partial S}{\partial y^a}=T_a(x)+
    o(y)=0\,,\qquad \mathbf{T}'_i=\frac{\partial S}{\partial x^i}={\partial_i T_a}y^a +
    o(y^2)=0\,.
\end{equation}
As is seen, the dynamics on $M$ do not decouple from those on the
augmented configuration space  $\mathcal{E}^\ast$ for arbitrary
boundary conditions of $y$'s, as opposite to (\ref{T-aug}).
Nonetheless, imposing zero boundary conditions on $y$'s, we can
satisfy the second group of equations in (\ref{varS}) with $y=0$
and $x$ is arbitrary.  Then the first group of equations reduces
to the original equations of motion on $M$.

An important observation on the action (\ref{Sxy}) is that it has
the form of local functional whenever the augmented constraints
(\ref{TT-aug}) are local\footnote{i.e., given by ordinary
functions of fields $(\phi^I, \bar\phi_J)$ and their derivatives
up to some finite order.}.  Indeed, the only place where
non-locality could emerge is the inversion of the augmented anchor
(\ref{V}). But, as is seen from (\ref{inv1}), the inversion
procedure, being performed perturbatively in $y$'s, does not spoil
locality. Therefore, the equivalent Lagrangian equations
(\ref{S'}) are local and so is the action functional (\ref{Shom}).

\subsection{Quantizing non-Lagrangian dynamics via augmentation}
As the augmented theory is always Lagrangian, its probability
amplitude  has the standard form
\begin{equation}\label{Psi}
    \Psi(x,y)=e^{\frac i\hbar S_\hbar(x, y)} \,,\qquad S_\hbar(x, y)=\sum_{n=0}^\infty
    \hbar^n S_n(x,y)\,.
\end{equation}
Here the leading term $S_0(x,y)$ is given by the classical action
(\ref{Sxy}) and the other terms can be regarded as quantum
corrections to the naive path-integral measure $dxdy$ on $\mathcal
E^\ast$. By definition (\ref{DQM}), the probability amplitude
(\ref{Psi}) is a unique solution to the Schwinger-Dyson equations
\begin{equation}\label{SD}
\hat{\widetilde {\mathbf T}}_I \Psi(x,y)=0
\end{equation}
associated with the (over)complete set of augmented constraints
(\ref{TT-aug}). Notice that the $\phi\bar\phi$-symbols of the
quantum constraint operators in (\ref{SD}) may defer from
(\ref{TT-aug}) by some quantum corrections in $\hbar$. These
corrections can be systematically derived by solving the quantum
master equation ${\hat \Omega}^2_{\mathrm{aug}}=0$ for the
augmented BRST operator.

What we are going to show in this section is that integrating the
amplitude (\ref{Psi}) of $y$'s, we get the solution to the
original Schwinger-Dyson equations (\ref{DQM}). In other words,
averaging the augmented probability amplitude (\ref{Psi}) over the
fibers of the vector bundle $\mathcal{E}^\ast\rightarrow M$ yields
the probability amplitude for the original (non-)Lagrangian
dynamics on $M$. We prove this statement under the following
technical assumptions:

\begin{itemize}
    \item [a)] The normal symbol of the augmented BRST operator $\hat\Omega_{\mathrm{aug}}$,
    which may differ from the classical BRST charge (\ref{augBRST}) by some
    quantum corrections, still obeys Rel. (\ref{cond}).
    \item [b)] Both the augmented and original  constraint operators are
    hermitian (w.r.t. the standard inner product associated with the translation-invariant integration measures $dxdy$ and $dx$ on $\mathcal{E}^\ast$ and $M$,
    respectively).
\end{itemize}
Note that the second condition follows from hermiticity
requirement for the BRST operator provided that the fields
$\bar\eta^I$ are chosen to be real, i.e., $(\bar\eta^I)^\ast
=\bar\eta^I$.

\begin{proposition}
Under the assumptions above,

(i) the physical observables of the original theory  are also
observables  of the augmented theory;

(ii)the functional
\begin{equation}\label{int1}
\psi(x)=\int dy \Psi(x,y)\,,
\end{equation}
where $\Psi(x,y)$ is the Feynman amplitude (\ref{Psi})  and the
integral is taken over all $y$'s satisfying zero boundary
conditions, obeys the original Schwinger-Dyson equations
(\ref{DQM}) in coordinate representation;

(iii) let $\mathcal{O}$ be the physical observable associated with
an on-shell gauge invariant function $O\in C^{\infty}(M)$ of the
original theory, then the quantum average (\ref{expect}) is given
by
\begin{equation}\label{int2}
\langle \mathcal {O}\rangle =(const)\int dydx\, {O}(x)\Psi(x,y)\,.
\end{equation}
\end{proposition}

\begin{rem}
The integrals (\ref{int1}) and (\ref{int2}), as they stand, are
well defined only for theories of type $(0,0)$. In presence of
gauge symmetries and/or Noether identities one should treat the
action $S_\hbar(x,y)$ within the usual BV quantization method.
This implies extension of the augmented configuration space
$\mathcal{E}^\ast$ by ghost fields and imposing gauge fixing
conditions that effectively reduces integration (\ref{int2}) to
the space of gauge orbits. It is a  perfectly standard technology
and we will not dwell on it here.
\end{rem}

\begin{proof}
Statement $(iii)$ is an immediate  consequence of $(i)$ and
$(ii)$.

We start with proving $(ii)$. By definition, the amplitude
$\Psi(x,y)$ is the matter state annihilated by the operators of
augmented constraints. In particular, it is annihilate by the
constraint operators that are extensions to $\mathcal{E}^\ast$ of
the original constraints (\ref{TRZ}). We have
\begin{equation}\label{label}
    \hat{\mathbf{\Theta}}_I \Psi (x,y)= 0\,,
\end{equation}
where the constraints $\mathbf{\Theta}_I =
(\widetilde{\mathbf{T}}_a, \widetilde{\mathbf{R}}_\alpha)$ are
defined by Rels. (\ref{TT-aug}) and (\ref{RR-aug}). The condition
(\ref{cond}), being imposed on the normal symbol of the augmented
BRST charge, suggests the following structure for the
$\phi\bar\phi$-symbols of the constraint operators:
\begin{equation}\label{prop}
{\mathbf{\Theta}}_I=\Theta_I(x,\bar x)+ o (\bar y).
\end{equation}
That is the difference $\mathbf{\Theta}_I-\Theta_I$ between the
original and augmented constraints is not only at least first
order in $y$'s it is also at least first order in $\bar y$'s. Now
multiplying (\ref{label}) on an arbitrary function of $x^i$ and
integrating the result over $\mathcal{E}^\ast$, we get
\begin{equation}\label{id}
\begin{array}{cc}
\displaystyle 0=\int dxdy\, \Phi(x)\hat{{\mathbf{\Theta}}}_I \Psi
(x,y)=\int dxdy\,
(\hat{{\mathbf{\Theta}}}{}^\dagger_I \Phi(x)) \Psi (x,y)\\[5mm]
\displaystyle =\int dx\,\hat{{\Theta}}{}^\dagger_I \Phi(x)\int
dy\, \Psi (x,y)=\int dx\, \Phi (x) \hat{{\Theta}}_I \int
dy\,\Psi(x,y)\,.
\end{array}
\end{equation}
Here we have used the hermiticity requirements
\begin{equation}
\hat{{\mathbf{\Theta}}}{}_I^\dagger=\hat{{\mathbf\Theta}}_I\,,\qquad\hat{\Theta}{}_I^\dagger=\hat{\Theta}_I\,,
\end{equation}
and  Rel. (\ref{prop}). Since the function $\Phi(x)$ is arbitrary,
the identity (\ref{id}) is equivalent to the desired one
\begin{equation}
 \hat{{\Theta}}_I\int dy\, \Psi(x,y)=0\,.
\end{equation}

Now, let us prove $(i)$. As we have already mentioned in
Sec.\ref{PO}, the space of physical observables is canonically
isomorphic to the space of on-shell invariant functions on
configuration space modulo trivial ones. In particular, a function
$O\in C^{\infty}(\mathcal{E}^\ast)$ gives rise to a BRST invariant
function on $\mathcal{N}_{\mathrm{aug}}$, i.e., an observable of
the augmented theory, iff
\begin{equation}\label{RR-inv}
    \mathbf{R}_\alpha O =\mathbf{W}_\alpha^a \mathbf{T}_a+\mathbf{W}_\alpha^i \mathbf{T}_i\,,
    \qquad \mathbf{R}_A O=\mathbf{W}_A^a\mathbf{T}_a +\mathbf{W}_A^i\mathbf{T}_i
\end{equation}
for some  $\mathbf{W}$'s. Here the generators of the augmented
gauge algebra may differ  from (\ref{RR-aug}) by quantum
corrections. Due to Proposition \ref{52} and our assumptions these
generators have the following structure:
\begin{equation}\label{}
    \mathbf{R}_\alpha =R_\alpha^i(x)\frac{\partial}{\partial
    x^i}+R^a_\alpha(x,y)\frac{\partial}{\partial y^a}\,,\qquad \mathbf{R}_A=R_A^a(x,y)\frac{\partial}{\partial
    y^a}\,.
\end{equation}
It remains to observe that for a $y$-independent function $O(x)$,
Eqs. (\ref{RR-inv}) reduce to the on-shell invariance condition
(\ref{on-sell-inv}),
\begin{equation}\label{}
    R_\alpha O =W_\alpha^a T_a\,.
\end{equation}
This completes the proof.
\end{proof}

\begin{example}
Consider a Lagrangian theory with action $S(x)$. In this case, the
dynamics bundle coincides with the cotangent bundle $T^\ast M$ of
the space of all histories. For simplicity sake assume that
$T^\ast M$ admits a flat connection. Given the canonical anchor
(\ref{V1}), the augmented constraints (\ref{TT-aug}) on $T M\oplus
T^\ast M\oplus T^\ast M$ read
\begin{equation}\label{}
    \widetilde{\mathbf{T}}_i=\partial_iS(x)+\bar x_i-\bar
    y_i\,,\qquad \widetilde{\mathbf{T}}'_i
    =\partial_iS(x+y)-\partial_iS(x)-\bar x_i\,.
\end{equation}
 The action of the augmented theory (\ref{Sxy}) takes the form
\begin{equation}\label{}
S(x,y)= S(x+y)-S(x)=y^i\partial_iS(x)+\frac 12
y^iy^j\partial_i\partial_jS(x)+\cdots\,.
\end{equation}
After normalization, the quantum average of a physical observable
$O(x)$ coincides with its usual value
\begin{equation}\label{}
    \langle O\rangle=\int_{TM} dxdy\, O(x)e^{\frac i\hbar
    S(x,y)}=(\mathrm{const})\int_M dx\, {O}(x)e^{-\frac i\hbar S(x)}\,.
\end{equation}
Of course, in the presence of gauge symmetries both these
integrals are to be understood as integrals  over the space of
gauge orbits rather than over $TM$ or $M$.

In this Lagrangian case, it is quite natural to interpret the
augmentation fields $y^i$ as the variations of the original fields
$x^i$  and this interpretation is automatically consistent with
the zero boundary conditions for $y$'s.
\end{example}

\section{Examples of quantizing non-Lagrangian field theories}

In this section, we demonstrate by examples what the Lagrange
anchor can look like in non-Lagrangian relativistic field theory
and how the general formalism described in previous sections works
in practice. As the examples we consider two illustrative
non-Lagrangian models: Maxwell electrodynamics with monopoles and
self-dual $p$-form fields.

The Maxwell equations are considered in terms of the strength
tensor and, in this formulation, they are not Lagrangian even
without magnetic currents. If a magnetic monopole was point-like
and satisfied the Dirac quantization condition, the theory would
admit an equivalent Lagrangian formulation in terms of vector
potential. We consider generic magnetic and electric sources, so
the theory does not have any Lagrangian reformulation, although it
still has a nontrivial Lagrange anchor, that is sufficient for a
consistent quantization of the model. The observation about the
structure of the Lagrange anchor might  also be instructive for
other non-Lagrangian field theories formulated in terms of
strength tensors.

Studying the second example, we reverse the order of exposing the
quantization procedure as compared to that described in Sections 4
and 5. Given the equations of motion for a non-Lagrangian field
theory in $d$ dimension and a compatible Lagrange anchor, the
general method allows one to equivalently reformulate this theory
as a topological Lagrangian field theory in $d+1$ dimensions, with
the original dynamics being localized at the boundary of this
$(d+1)$-dimensional  space-time. This also allows for a reverse
consideration: one can start with an appropriate
$(d+1)$-dimensional, topological Lagrangian theory and then
identify ``original'' $d$-dimensional field equations  and a
Lagrange anchor in the action of the topological theory. In
practice, this can be an instructive scheme for identifying those
non-Lagrangian models that admit Poincar\'e covariant Lagrange
anchors. To exemplify this idea, we take the Chern-Simons theory
in $4n+3$ dimensions and reinterpret it as resulting from some
quantization of self-dual $(2n+1)$-form fields in $4n+2$
dimensions.

\subsection{Maxwell electrodynamics with monopoles}

Consider the Maxwell equations with electric and magnetic
currents:
\begin{equation}\label{ED}
    d^\dagger \widetilde{F}=I\,,\qquad d^\dagger F=J\,.
\end{equation}
Here $F$ is the strength of the electromagnetic field, whose Hodge
dual is denoted by $\widetilde{F}=\ast F$, $J$ and $I$ are the
electric and magnetic currents, respectively, and $d^\dagger =\ast
d\ast$ is the adjoint exterior differential. As a consequence of
Eqs.(\ref{ED}), the currents $J$ and $I$ are conserved,
\begin{equation}\label{}
d^\dagger I=0\,,\qquad d^\dagger J=0\,.
\end{equation}

Clearly, Equations (\ref{ED}), as they stand, are not Lagrangian,
even if we set $I=0$. (The number of equations is less than the
number of fields).

Let us introduce the source $P$ which is canonically conjugate to
the field $F$. With the field $F$, being a 2-form on the
space-time manifold, the source $P$ is a bivector field on the
same manifold. The canonical symplectic structure on the cotangent
bundle of the space of all histories is given by
\begin{equation}\label{}
\omega =\int \delta \widetilde{F} \wedge \delta P'\,,
\end{equation}
where the 2-form $P'$ is obtained from $P$ by lowering the upper
indices with the space-time metric.

Consider now the following set of first-class constraints on the
phase space of fields and sources:
\begin{equation}\label{constr}
\begin{array}{c}
    T^1=d^\dagger\widetilde{F}-I\approx 0\,,\qquad
    T^2=d^\dagger(F+P')-J\approx 0\,,\\[5mm]
    \{T^{a},T^{b}\}=0\,,\qquad a,b=1,2\,.
\end{array}
\end{equation}
These constraints are obtained from (\ref{ED}) by adding the
momentum depending term $d^\dagger P'$ to the second group of
equations. It is the term that defines the canonical Lagrange
anchor for the  Maxwell electrodynamics \cite{KLS}. Observe that
the anchor is regular but not complete (see Definition 2.3). The
physical meaning of this incompleteness can be understood in the
following way. Let $I=0$, then the first group of equations
(\ref{ED}) expresses the closedness condition for the strength
form $F$. The absence of momentum contributions to the
corresponding constraints $T^1\approx 0$ implies that we consider
these equations as being pure non-Lagrangian in the sense of
Theorem 2.5. Hence, no quantum fluctuations violate the closedness
condition $dF=0$, that guarantees  the existence of a local gauge
potential $A=d^{-1}F$ both at classical and quantum levels.

Notice that the constraints (\ref{constr}) are linearly
dependent, $ d^\dagger T^a= 0$, while the classical equations of
motion (\ref{ED}) are not gauge invariant. Thus, according to
Definition 2.1, we have a theory of type $(0,1)$.

Upon canonical quantization, the constraints (\ref{constr}) turn
into the following Schwinger-Dyson operators:
\begin{equation}\label{}
\hat{T}^1=d^\dagger\widetilde{F}-I\,,\qquad\hat{T}^2 =
d^\dagger\left(F-i\hbar \frac{\delta}{\delta F'}\right)-J\,,
\end{equation}
$F'$ being the contravariant  strength tensor of electromagnetic
field. The corresponding Schwinger-Dyson equation for the
probability amplitude
\begin{equation}\label{TPSi}
\hat{T}^a\Psi[F]=0
\end{equation}
is satisfied  by
\begin{equation}\label{max-amp}
    \Psi[F]=\Delta[T^1]e^{\frac i\hbar S[F]}\,.
\end{equation}
Here
\begin{equation}\label{}
    \displaystyle S[F]=\int \frac12 Gd \widetilde{F}\wedge \ast d {\widetilde{F}}-G  \widetilde{F}\wedge
    d J\,,\qquad
    \displaystyle \Delta [T^1]=\int DC \delta[T^1-dC]\,,
\end{equation}
$C$ is an auxiliary  0-form,  and $G$ is the inverse of the
Laplace operator $\Box=dd^\dagger+d^\dagger d$. One can easily see
that the distribution  $\Delta[T^1]$, considered as the functional
of $F$, is supported at the points where $T^1[F]=0$ so that
$T^1\Delta[T^1]=0$. (A naive solution to the last equation, namely
$\Delta[T^1]=\delta[T^1]$, is ill defined because of linear
dependence of the constraints $T^1$.) Notice that the amplitude
(\ref{max-amp}) is non-Feynman: it is a nearly everywhere
vanishing  distribution on the configuration space of fields
rather than a smooth, complex-valued function with absolute value
1. This fact is a direct consequence of incompleteness of the
Lagrange anchor discussed above.

Passing to the momentum representation, we get the generating
functional of Green's functions
\begin{equation}\label{}
\begin{array}{c}
\displaystyle    Z[P]= \int DF \Psi[F]e^{\frac i\hbar \int
P'\wedge
    \widetilde{F}}= e^{\frac i\hbar W[P]}\,,\\[5mm]
\displaystyle    W[P]=\int \frac 12 Gd^\dagger P' \wedge \ast
d^\dagger P'-\bar F \wedge \ast P'\,,
    \end{array}
\end{equation}
where
\begin{equation}\label{}
\bar F=dGJ+\ast dGI
\end{equation}
is the mean electromagnetic field produced by the sources $I$ and
$J$. As with any free theory, the mean field $\bar F$ satisfies
the classical equations of motion (\ref{ED}). One can also see
that the propagator $\langle F(x)F(x')\rangle$ for the field $F$
coincides with corresponding expression  $\langle
dA(x)dA(x')\rangle$ in the Maxwell electrodynamics with action
$S[A]=\frac 12\int dA\wedge\ast dA$.

The probability amplitude (\ref{max-amp}) can also be arrived at
by applying the augmentation method. By definition, the
augmentation fields are the sections of the bundle which is  dual
to the dynamics bundle of the theory. So we introduce the 1-forms
$B^a$, $a=1,2$; the pairing between the equations of motion and
augmentation fields is given by the integral $\int T^a\wedge \ast
B_a$. Since the constraints (\ref{constr}) are linear in fields
and momenta, the action $S[F,B]$ of the augmented theory is at
most  quadratic in $F$ and $B^a$. Specializing  the general
formulas (\ref{Sxy}) and (\ref{VV}) to the case at hand, we find
\begin{equation}\label{SFB}
    S[F,B]=\int (d^\dagger \widetilde{F}-I)\wedge \ast B_1 + (d^\dagger F-J)\wedge \ast B_2  +
    \frac12dB_2\wedge \ast dB_2.
\end{equation}
The Noether identities between the original equations of motion
(\ref{ED}) give rise to the gauge invariance of the action
(\ref{SFB}):
\begin{equation}\label{}
    \delta_\varepsilon B_a=d\varepsilon_a\,,\qquad a=1,2\,.
\end{equation}
We can fix this arbitrariness by imposing the Lorentz gauges
$d^\dagger B^a=0$ on the augmentation fields and adding these
constraints to the action (\ref{SFB}) with the Lagrange
multipliers $C^a$. Then the gauge-fixed  action reads
\begin{equation}\label{SGF}
    S_{\mathrm{gf}}[F,B,C] =S[F,B]+\int  d^\dagger B_a\wedge \ast
    C^a\,.
\end{equation}
According to Proposition 5.4, the (non-Feynman) probability
amplitude (\ref{max-amp}) admits the following path-integral
representation in terms of the local action (\ref{SGF}):
\begin{equation}\label{psi-aug}
    \Psi[F]=(const)\int DBDC e^{\frac i\hbar S_{\mathrm{gf}}[F,B,C]}
\end{equation}
Of course, in the case under consideration, one can verify the
last equality directly, either by calculating the Gauss integrals
over $B$'s and $C$'s or substituting (\ref{psi-aug}) into the
Schwinger-Dyson equation (\ref{TPSi}) and differentiating under
the integral sign.

Given the probability amplitude (\ref{max-amp}), the quantum
average of a physical observable ${O}$ is defined by the path
integral \begin{equation}\label{max-int}
    \langle{O}\rangle=\int DF {O}[F]\Psi[F]\,.
\end{equation}
In case $I=0$, one can solve  the constraint $T^2=d^\dagger
\widetilde{F}\approx 0$ in terms of the gauge potential $A$
obeying the Lorentz gauge-fixing condition,
\begin{equation}\label{}
    F=dA\,,\qquad d^\dagger A=0\,,
\end{equation}
and integrate the pre-exponential $\Delta$-functional in
(\ref{max-amp}) as
\begin{equation}\label{}
    DCDF\Delta[d^\dagger \widetilde{F}+dC]\quad\rightarrow \quad DA\delta[d^\dagger A]\,.
\end{equation}
Then  the integral (\ref{max-int}) takes the form
\begin{equation}\label{}
  \langle  {O}\rangle =\int DA
  {O}[F(A)]\Psi[A]\,,
\end{equation}
where
\begin{equation}\label{}
    \Psi[A]=\delta[d^\dagger A]\exp\frac i\hbar \int\frac12 dA\wedge\ast dA
    +A\wedge \ast J
\end{equation}
is nothing but the usual  probability amplitude for the
electromagnetic field subject to the Lorentz gauge.

\subsection{Self-dual $p$-form fields} It has long been known
that the quantization  of chiral bosons in $(4n+2)$-dimensional
space-time is closely related with the quantization of
Chern-Simons theory in the space-time with one more dimension
\cite{Witten}. Roughly speaking, a physical wave-function of
Chern-Simons fields on a $(4n+3)$-dimensional manifold
$\mathcal{M}$ can be treated as a probability amplitude (or
partition function) for the self-dual fields living on the
boundary of $ \mathcal{M}$. For a recent discussion of the
relationship between self-dual fields and the Chern-Simons theory
we refer the reader to \cite{BM}. Below we justify and reinterpret
this \textit{ad hoc} quantization technique within the general
method of Sections 4 and 5.

 Our starting point is the
Chern-Simons  action for the $(2n+1)$-form field $F\in
\Lambda^{2n+1}(\mathcal{M})$ on a $(4n+3)$-dimensional manifold
$\mathcal{M}$
\begin{equation}\label{CS}
    S=-\frac12\int_\mathcal{M}F\wedge DF\,,
\end{equation}
$D$ being the exterior differential on $\mathcal{M}$. Assume that
$\mathcal{M}=M\times I$, where $I=[0,1]\subset \mathbb{R}$  and
$M$ is a compact $(4n+2)$-dimensional manifold without boundary.
Then $\partial \mathcal{M}=M\cup M$.

Using the product structure of the manifold $\mathcal{M}$, one can
globally decompose the field $F$ and the operator $D$ as
\begin{equation}\label{}
    F=H+B\wedge dt\,, \qquad D=dt\wedge\partial_t+d\,,
\end{equation}
Here $H\in \Lambda^{2n+1}(M)$ and $B\in \Lambda^{2n}(M)$ are the
one-parameter families of differential forms labelled by $t\in
[0,1]$ and $d$ is the exterior differential on $M$. In this
notation, the action (\ref{CS})  takes a simple Hamiltonian form
(\ref{top}), if one identifies $t$ with  evolution parameter:
\begin{equation}\label{HamAct}
    S=\int_I dt\int_{M}\left(\frac12H\wedge \dot H-B\wedge
    dH\right)\,.
\end{equation}
The fist term in (\ref{HamAct}) defines (and is defined by) a
symplectic structure on $\Lambda^{2n+1}(M)$; the corresponding
symplectic 2-form reads
\begin{equation}\label{omega-H}
    \omega=\int_M\delta H\wedge\delta H\,.
\end{equation}
The field $B$ plays the role of the Lagrange multiplier to the
first-class constraints
\begin{equation}\label{TdH}
{T}=dH\approx 0\,.
\end{equation}
It is convenient to treat $T$ as a linear functional (de Rham's
flux) on the space of $2n$-forms:
\begin{equation}\label{}
    T[\alpha]=\int_M\alpha \wedge dH\,,\qquad \forall\alpha\in \Lambda^{2n}(M)\,.
\end{equation}
Then one can easily check that the constraints (\ref{TdH}) have
vanishing Poisson brackets,
\begin{equation}\label{}
    \{T[\alpha], T[\beta]\} =\int_M d\alpha\wedge d\beta =0\,,\qquad
    \forall \alpha,\beta\in \Lambda^{2n}(M).
\end{equation}

Since $d^2=0$, these constraints are reducible, $dT\equiv 0$, and
one can further deduce that the order of reducibility is $2n$.

The Hamiltonian reduction by the first-class constraints
(\ref{TdH}) leads to a finite dimensional phase space.  We have a
rather explicit description of the reduced phase space due to the
Hodge decomposition
\begin{equation}\label{}
 \Lambda^{2n+1} (M)=d\Lambda^{2n}(M)\oplus d^\dagger
    \Lambda^{2n+2}(M)\oplus \Lambda^{2n+1}_{H}(M)\,.
\end{equation}
Here $d^\dagger : \Lambda^m(M)\rightarrow \Lambda^{m-1}(M)$ is the
adjoint differential constructed by some  Riemannian metric on
$M$, and $\Lambda_H^{2n+1}(M)$ is the subspace of harmonic forms.
According to the Hodge theory, the space $\Lambda^{2n+1}_H (M)$ is
naturally isomorphic to the de Rham cohomology group
$H^{2n+1}(M)$. The first-class constraints  (\ref{TdH}) single out
the coisotropic subspace of $d$-closed forms, whose complementary
isotropic subspace is given
 by the $d^\dagger$-exact forms. Notice that both these subspaces
are Lagrangian iff $H^{2n+1}(M)=0$.

The Hamiltonian flux generated by the first-class constraints
changes any $(2n+1)$-form $H$ by an exact one:
\begin{equation}\label{}
    \delta_\varepsilon H=\{H,T[\varepsilon]\}=d\varepsilon\,.
\end{equation}
Taking the quotient of $d$-closed $(2n+1)$-forms by $d$-exact
ones, we obtain the physical phase space of the model, which is
apparently  isomorphic to the (finite dimensional) subspace of
harmonic forms on $M$:
\begin{equation}\label{}
    \Lambda^{2n+1}(M)//T \simeq \Lambda ^{2n+1}_H(M)\simeq
    H^{2n+1}(M)\,.
\end{equation}
We are lead to conclude that the model under consideration is not
topological unless $H^{2n+1}(M)=0$. (We define a topological
theory  as a theory without physical degrees of freedom.) To get
rid of the physical modes and obtain a pure topological model we
can restrict the dynamics on the affine subspace
$\Lambda_\alpha^{2n+1}(M)\subset \Lambda ^{2n+1}(M)$ constituted
by the forms $H=H_\alpha +H_0$, where $H_0\in
d\Lambda^{2n}(M)\oplus d^\dagger\Lambda^{2n+2}(M)$ and $H_\alpha$
is a time-independent harmonic $(2n+1)$-form representing the de
Rham class $\alpha=[H_\alpha]\in H^{2n+1}(M)$. This restriction is
compatible with dynamics. Indeed, the equations of motion
following from the Hamiltonian action (\ref{HamAct}) read
\begin{equation}\label{HamEqs}
    \dot H = dB\,,\qquad dH=0\,.
\end{equation}
So, the de Rham class $[H]$ of the closed form $H$ does not change
with time and can thus  be regarded as a (topological) integral of
motion. Moreover, the embedding $\Lambda_\alpha^{2n+1}(M)\subset
\Lambda^{2n+1}(M)$ is symplectic, i.e., the restriction of the
$2$-form (\ref{omega-H}) to $\Lambda_\alpha^{2n+1}(M)$ is
nondegenerate.

To further proceed with the interpretation and quantization of the
Chern-Simons theory on the product manifold $\mathcal{M}=M\times
I$, let us endow $M$ with a Lorentzian metric. (A necessary and
sufficient condition for such a metric to exist is that the Euler
characteristic $\chi(M)$ be zero.) Then, the corresponding Hodge
operator $\ast$ squares to $+1$ on the middle forms so that any
$(2n+1)$-form $H$ admits a unique decomposition in the sum of its
self-dual and anti-self-dual parts:
\begin{equation}\label{}
    H=H^++H^-\,,\qquad \ast H^{\pm}=\pm H^{\pm}\,.
\end{equation}
Since $\omega(\delta H^{\pm}, \delta H^\pm)=0$, we have a natural
polarization of the phase space $\Lambda^{2n+1}(M)$ given by the
two complementary Lagrangian subspaces of self- and anti-self-dual
forms:
\begin{equation}\label{polarization}
    \Lambda^{2n+1}(M)=\Lambda^{2n+1}_+(M)\oplus\Lambda^{2n+1}_-(M)\,.
\end{equation}

Let us regard the fields $H^-$ as the ``momentum coordinates''
canonically conjugate to the ``position coordinates'' $H^+$ and
rewrite the Hamiltonian action (\ref{HamAct}) as
\begin{equation}\label{SL}
    S=\int_I dt\int_{M}H^-\wedge \dot H^+-B\wedge d(H^++H^-)
\end{equation}
Upon restriction to $\Lambda_\alpha^{2n+1}(M)$ this action
describes a topological field theory and its form is identical to
the form of the topological action (\ref{top}). Recall that the
latter was constructed on general grounds starting from some
classical (not necessarily Lagrangian) equations of motion
supplemented with an appropriate Lagrange structure. Proceeding
now backward, we can readily reinterpret the topological model
(\ref{SL})  in terms of non-Lagrangian dynamics on $M$.

Namely, the classical equations of motion are to be identified
with the momentum independent terms in the Hamiltonian constraints
\begin{equation}\label{dHH}
    T=d(H^++H^-)\approx 0  \,.
\end{equation}
With our choice of the phase-space polarization this yields the
closedness condition for the self-dual form $H^+$,
\begin{equation}\label{dH+}
  dH^+=0\,.
\end{equation}
Similar to the Hamiltonian constraints (\ref{dHH}), these
equations are $2n$-times reducible, and hence they define a
regular gauge theory of type $(0,2n)$ (although there is no gauge
invariance in the usual sense). A non-Lagrangian nature of
equations (\ref{dH+}) was discussed at length in \cite{MS}.

The second term in the Hamiltonian constraints (\ref{dHH}), namely
$dH^-$, is linear in momenta and should be identified with the
Lagrange anchor. Zero boundary conditions (\ref{bcon}) on the
momenta
\begin{equation}\label{}
H^-|_{t=0,1}=0 \,
\end{equation}
ensure the equivalence of the classical dynamics (\ref{HamEqs})
and (\ref{dH+}), where both $H$ and $H^+$ belong to
$\Lambda_\alpha^{2n+1}(M)$. Fixing the de Rham class
$[H^+]=\alpha$ of a solution to equation  (\ref{dH+}) is quite
similar to fixing the boundary conditions for a field theory on a
bounded space-time domain.

Quantizing now the Hamiltonian constraints (\ref{dHH}) in the
coordinate representation, we get the following Schwinger-Dyson
operator:
 \begin{equation}\label{}
    \hat{T}=d\left(H^+-i\hbar\frac\delta{\delta H'{}^+}\right)\,.
\end{equation}
Here $H'{}^+$ is the bivector on $M$ obtained from $H^+$ by rising
indices with the help of the Lorentzian metric. The probability
amplitude  on the configuration space
$\Lambda_\alpha^{2n+1}(M)\cap \Lambda_+^{2n+1}(M)$ is determined
by the equation
\begin{equation}\label{TH+}
    \hat{T}\Psi[H^+]=0\,.
\end{equation}

We use the augmentation method to write down an explicit
path-integral representation for $\Psi[H^+]$. To this end, we
introduce the augmentation fields $C\in \Lambda^{2n}(M)$ whose
configuration space is dual to the linear space
$\Lambda^{2n+2}(M)$ of the field equations (\ref{dH+}), and apply
the general formulas (\ref{Sxy}), (\ref{VV}), and (\ref{Psi}) to
construct the Feynman probability amplitude of the augmented
theory. The result is almost obvious:
\begin{equation}\label{CH+}
\begin{array}{c}
    \Psi [C, H^+]=e^{\frac i\hbar S[C,H^+]}\,,\\[5mm]
    \displaystyle   S[C,H^+]=\int -\frac12 dC\wedge \ast dC+H^+\wedge
    dC\,.
    \end{array}
\end{equation}
The corresponding equations of motion  read
\begin{equation}\label{}
    dH^+=0\,,\qquad (dC)^-=0\,.
\end{equation}
The augmented theory is seen to describe the pair of self-dual
fields: one in terms of the ``strength tensor'' $H^+$ and another
one in terms of the ``gauge potential'' $C$. Notice that the
Noether identities for the non-Lagrangian equations of motion
(\ref{dH+}), i.e., $d(dH^+)\equiv 0$,  reincarnate as the gauge
transformations of the augmentation fields:
\begin{equation}\label{GF}
    C\rightarrow C'=C+d A\,,\qquad \forall A\in \Lambda^{2n-1}(M)\,.
\end{equation}

Integrating formally the amplitude (\ref{CH+}) over the fields
$C$, we obtain the probability amplitude for the self-dual field
$H^+$,
\begin{equation}\label{GI}
    \Psi[H^+]=\int DC \Psi[C,H^+]\,.
\end{equation}
It is now just a matter of differentiating under the integral sign
to show that the amplitude (\ref{GI}) does obey the
Schwinger-Dyson equation (\ref{TH+}). We have
\begin{equation}\label{}
\hat{T}\Psi[H^+] = \int DC \hat{T}e^{\frac i\hbar S[C,H^+]}=i\hbar
\ast \int DC\frac{\delta}{\delta C}e^{\frac i\hbar S[C,H^+]}=0\,.
\end{equation}
A more rigor treatment of the (divergent) Gaussian integral
(\ref{GI}) implies fixing the gauge freedom (\ref{GF}) by the BV
method for reducible gauge-algebra generators \cite{BV},
\cite{HT}.

\end{document}